\documentclass[12pt,a4paper]{article}
\usepackage[utf8]{inputenc}
\usepackage{amsmath}
\usepackage{amsfonts}
\usepackage{amssymb}
\usepackage{graphicx}
\usepackage[left=2cm,right=2cm,top=2cm,bottom=2cm]{geometry}
\usepackage[varg]{txfonts}
\usepackage{bm}
\usepackage{algorithm2e}
\usepackage{xcolor}

\title{Modelling the Proliferation of Terrorism Via Diffusion and Contagion}
\author{Gentry White and Fabrizio Ruggeri and Michael D. Porter}
\date{}
\begin{document}
\maketitle
\begin{abstract}
The spread of terrorism is a serious concern in national and international security, as its spread is seen as an existential threat to Western liberal democracies. Understanding and effectively modelling the spread of terrorism provides useful insight into formulating effective responses. 
A mathematical model capturing the theoretical constructs of contagion and diffusion is constructed for explaining the spread of terrorist activity and used to analyse data from the Global Terrorism Database from 2000--2016 for Afghanistan, Iraq, and Israel.
Results show that the model identifies patterns in the diffusion and contagion processes that align with and provide insight into contemporary events.  
\end{abstract}
%\keywords{terrorism, contagion, diffusion, self-exciting process}
\section{Introduction}
%%  The Introduction

Terrorism is a complicated social phenomenon, and assessing the effectiveness of measures to prevent the proliferation or increase in the rate of terrorist events relies on a model that incorporates a qualitative and theoretical understanding of the process while maintaining fidelity to the data, hence accurately reflecting the effects of counter-measures. This paper presents such a model, it incorporates an accepted theoretical framework and refined elements of previous quantitative models and is parametrised to allow the extraction of meaningful insights from the results.  

There are two aspects to acts of terrorism\footnote{The 
definition of terrorism is contested (see \cite{Schmid:Jongman:2005,Hoffman:2006}, for an entr\'e into this discussion), in this paper 
the term ``terrorist'' and ``terrorism'' refer only to the fact that the 
events analysed in this paper are drawn from the Global Terrorism 
Database \cite{GTD:2012}.} that exacerbate their impact.
The first is their relative rarity (in most contexts), the second is that the clustering of events in time \cite{Li:Thompson:1975,Midlarsky:1978,Heyman:Micklous:1980,Midlarsky:etal:1980,Holden:1986,Telesca:Lovallo:2006,Townsley:2008,Lafree:etal:2009,loepucl1354681} creates the impression of a sudden increase in the rate of terrorism.  It is the uncertainty surrounding the permanence of this increase that causes additional anxiety \cite{Thorton:1964,Hutchinson:1972,Crenshaw:1981,Crenshaw:1986,Bandura:1990}.  Determining whether the occurrence of a terrorist event is a transient anomaly or a signal of a change in the rate of events is of great interest and identifying the circumstances around these two possibilities is a feature of the model presented in this paper. Theoretical models posit the mechanisms of contagion and diffusion as explanations for these patterns of clustering and erratic activity. Each of these mechanisms has different implications for identifying whether events are part of a persistent increase in the rate of events or are part of a transient anomaly.  
%\mdp{(This last sentence isn't clear. events part of increase of rate or anomaly? Don't both mechanisms cause increase in rate? )}\\
%\gaw{(See my change, I added the word persistent to contrast the term transient.)}
%These two mechanisms emerge as explanation for the spread of 
%terrorism almost simultaneously, though it is \cite{Midlarsky:1978} 
%that notes that the spread of urban riots in the US during the 1960's, 
%was due to both contagion and diffusion.   

%\mdp{(Add clarification to this paragraph)}\\
%\gaw{I have tried to clarify things a bit, see my comments}

Sociological theories to explain the proliferation of aberrant behaviour through contagion or diffusion\footnote{The distinction between diffusion and contagion can be difficult to discern in the sociological literature. In some cases the terms seem to be used interchangeably. This is specifically acknowledged in \cite{Midlarsky:1978} which concludes by drawing a distinction between the two mechanisms.} begin with \cite{LeBon:1896}, and appear in discussions of political violence and civil upheaval by the mid-twentieth century \cite{Hopper:1950},  enduring as a viable theoretical model to explain the proliferation of terrorism both spatially and temporally \cite{Midlarsky:1978,Pitcher:etal:1978,Midlarsky:etal:1980,Hamilton:Hamilton:1983,Hill:Rothchild:1986,Holden:1986,Weinmann:etal:1988,Gurr:1993,Brynjar:Skjolberg:2000,Buhaug:Gleditsch:2008,Braithwaite:2010,Cliff:First:2013}.  The mathematical and statistical concepts of a “contagious” process are similarly well established \cite{Greenwood:Yule:1920,Neyman:1939,Feller:1943}, as are the concepts of diffusion \cite{Fick:1855,Einstein:1905, Onsager:1931,Teorell:1936,Philibert:2005}. The application of a model based approach to describing the dynamics of terrorism emerges in the social sciences with \cite{Coleman:1964}, followed by the development of theoretically based models \cite{Pitcher:etal:1978,Diekmann:1979,Hamilton:Hamilton:1981, Hamilton:Hamilton:1983}, building on theoretical constructs for social contagion, diffusion, and group learning. 
%\mdp{(something missing here)}\gaw{(see my changes)}
These models demonstrate a sophisticated understanding of the theories and mathematical models but focus on a simple mathematical mechanisms to describe the proliferation of violence.
%These theoretically based models while sophisticated and insightful in some cases lack desirable mathematical characteristics such as stationarity, making them technically incompatable with the theoretical definition of contagion as a purely transient, i.e. stationary process.   
Contemporary to these models the Hawkes self-exciting process model \cite{Hawkes:1971a,hawkes71b} and its cluster process interpretation \cite{hawkes&oakes:1974} are put forth as theoretical constructs with limited demonstrations of their application. What is important about these models is that they demonstrate a higher degree of mathematical sophistication than the ad hoc or heuristic models of \cite{Pitcher:etal:1978,Diekmann:1979,Hamilton:Hamilton:1981, Hamilton:Hamilton:1983}, having nice mathematical properties (e.g. stationarity), while accurately reflecting the theoretical understanding of the role behavioural contagion plays in the rate of terrorist incidents.

While the distinction between diffusion and contagion as mechanisms for the proliferation of terrorism is not always clear
\cite{Midlarsky:1978} identifies diffusion as the increase in the rate of events due to non-terrorist events that elicit a collective reaction among individuals or groups in a population without 
observation or communication of others' behaviour.  This is distinct from contagion as it contains no element of imitation or the modelling of behaviour, rather it implies a shift in an individual's or groups propensity to engage in terrorism as a reaction to external events.  If these shifts in patterns of behaviour are enduring, then the mean of the process can shift, implying the possibility that diffusion can be a non-stationary process.   Originally, \cite{Midlarsky:1978} applied this theoretical model, in conjunction with contagion, to explain the proliferation of riots in US cities during the late 1960's.  But it is reasonable to extend this to terrorism more broadly, including explanations for groups as well as individuals, as diffusion in \cite{Midlarsky:1978} is a reaction to what \cite{Crenshaw:1981} refers to as precipitants, or precipitating events, identified as an element in the proliferation of terrorism.
The resulting theoretical model states that terrorism proliferates through two mechanisms: contagion, which is a function of the influence that past terrorist events has on the future event rate, and diffusion, which is a function of exogenous events or processes.
This serves as the basis for the model presented in this paper, which describes
the combined effects of diffusion and contagion as a convolution of two process: a non-homogeneous Poisson process for diffusion (as originally proposed in \cite{Midlarsky:1970}) and a negative-binomial Hawkes self-exciting process for contagion. Results show that under the cluster process representation the model can identify distinct behaviours in the two processes that may be attributable to observable differences in the contexts and populations, providing useful insight into the actual phenomenon and data. 

The balance of this paper is structured as follows: Section 2 derives a Bayesian hierarchical model, incorporating diffusion and contagion components, for the convolution of a Poisson and negative-binomial processes. Section 3 provides the computational details of the model.
Section 4 presents the results of applying the model to multiple data sets. Section 5 concludes with a discussion of the results and their interpretation. 
\section{Model\label{Model}}
%%  The Model

The daily number %rate 
of terrorist events can be described as
%as an non-homogeneous Poisson process based on the Hawkes self-exciting process model \cite{Hawkes:1971a,hawkes71b,Porter:White:2012,White:etal:2012b,White:Porter:2013}.  The Hawkes model can be written 
%as the convolution of two Poisson processes, i.e. the total observed number of events is
the sum of %the number of 
events from two processes, a diffusion process and a contagion process.  
%The model for the convolution of the two process is based on the cluster process implementation of the Hawkes process model \cite{hawkes&oakes:1974}.  
As proposed in \cite{Midlarsky:1978}, 
a Poisson distribution is used for the diffusion process and a 
negative-binomial distribution is used for the contagion process 
resulting in a convolution model for the total number of daily 
events.
A closed form for the likelihood of a Poisson-Negative-Binomial convolution is not available.  This is addressed by using a Bayesian hierarchical model that facilitates computation by exploiting the relationship between the Poisson, gamma, and negative-binomial distributions. 

\subsection{A Hierarchical Model for the Convolution of Diffusion and Contagion Processes}
Evaluating a model for the convolution of two Poisson processes is straightforward, as it is simply another Poisson process with an explicit and easily evaluated likelihood.  
However, preliminary data analysis and results from \cite{White:etal:2012b} confirm 
%\mdp{(didn't our JQC paper also show this?)}\gaw{(see my change)} 
that data from the contagion process are better fit using a negative binomial distribution. 
The resulting Poisson-Negative-Binomial convolution model has no closed form likelihood function, complicating analysis.  This complication is addressed by noting that the total number of events on day $t$, $Y_t =Y_t^d + Y_t^c$ is the sum of the number of events from the diffusion and contagion processes, respectively, and specifying a hierarchical convolution model, %The convolution model is written using a hierarchical model.  
%\begin{eqnarray}
%\label{eq:Poisson}
%Y_t^d&\sim&Pois\left(\lambda_t^d\right)\\
%\label{eq:nb}
%Y_t^c&\sim&Neg-Binom\left(\mu_t^c,\sigma^2\right)
%\end{eqnarray}
%which doesn't have a convenient form for the likelihood of $Y_t=Y_t^d+Y_t^c$, but can be written as a hierarchical model
\begin{eqnarray}
\label{eq:Poisson2}
Y_t^d&\sim&Pois\left(\lambda_t^d\right)\\
\label{eq:PossionC}
Y_t^c|\lambda_t^c&\sim&Pois\left(\lambda_t^c\right)\\
\label{eq:gamma}
\lambda_t^c&\sim&Ga\left(\sigma^2, \frac{\mu_t^c}{\sigma^2}\right).
\end{eqnarray}
Note that \eqref{eq:gamma} is parametrised so that $E(\lambda_t^c)=\mu_t^c$, and that multiplying \eqref{eq:PossionC} and \eqref{eq:gamma} and integrating out $\lambda_t^c$ yields
\begin{equation}
\label{eq:nb}
Y_t^c \sim Neg-Binom\left(\mu_t^c,\sigma^2\right),
\end{equation}
where $\mu_t^c$ is parametrised as the expected value. %for the contagion process.

%The observed data are $Y_t=Y_t^d+Y_t^c$, and i
If the partitioning of $Y_t$ into  $Y_t^d$ and $Y_t^c$ were known, then the parameters for the two processes could be estimated using the \eqref{eq:Poisson2} and \eqref{eq:nb}.
%In either case evaluating these models, without an explicit form of the convolution likelihood, require knowing how the data are partitioned between the two processes.   
If 
$Y_t|\lambda_t^c\sim Pois(\lambda_t^d+\lambda_t^c)$ as implied by \eqref{eq:Poisson2} and \eqref{eq:PossionC} then from \cite{D/VJ:03} 
\begin{eqnarray}
\label{eq:binom}
Y_t^d|Y_t,\lambda_t^c\sim Binom\left(
Y_t,\frac{\lambda_t^d}{\lambda_t^d+\lambda_t^c}\right),
\end{eqnarray}
%The fact that $Y_t^c=Y_t-Y_t^d$ 
which can be used to stochastically attribute events to diffusion and contagion. 
%partition the events into each process. 
These results can be used to construct a hierarchical model and an MCMC scheme that does not require explicit evaluation of the likelihood for $Y_t$, but instead uses the partitioned data with  \eqref{eq:Poisson2} and \eqref{eq:nb} to draw samples from the posterior distributions of their parameters.  
%This is accomplished by partitioning the data at each iteration into $Y_t^d$ and $Y_t^c$ and  updating parameters using.  
Details of the implementation are in Section \ref{comp}.  

\subsection{Incorporating Exogenous and Endogenous Effects}
\label{IncEffects}
%In keeping with the definition of endogenous factors for contagion processes in \cite{Hamilton:Hamilton:1981,Hamilton:Hamilton:1983} and the conceptual model of preconditions and precipitants in \cite{Crenshaw:1981}, 
The model presented in \eqref{eq:Poisson2}--\eqref{eq:nb} can be extended to allow variation in diffusion and contagion rates due to other factors. \cite{Hamilton:Hamilton:1981} divides these factors into endogenous, associated with the contagion process (e.g., timing, casualties, attack characteristics), and exogenous effects that influence the diffusion process (e.g., what \cite{Crenshaw:1981} refers to as precipitating events or preconditions);  
%external to the contagion process (e.g., measures of economic growth and stability, and measures of civil liberties), 
%In this paper these definitions are refined to define endogenous effects as characteristics of prior terrorist events themselves , and exogenous effects as factors separate from the terrorists events (i.e. , including what \cite{Crenshaw:1981} identify as preconditions and precipitants, or other societal measures). 
thus exogenous effects are present in the diffusion process and endogenous effects in the contagion process. 
%and should be parametrised in the model to reflect this. 
%Using the notation in \eqref{eq:hawkes_intensity} this can be simplified by stating that endogenous effects are indexed by the time variable $s$ and exogenous effects are indexed by the time variable $t$; thus endogenous effects are located with in the summation and exogenous effects outside the summation.  
%\subsection{Incorporating Endogenous and Exogenous Variables}
%Exogenous factors are incorporated into the diffusion process by defining 

Following the convention for generalised linear models, the rate of the diffusion process is therefore modelled 
\begin{eqnarray}
\label{eq:Poisson3}
\lambda_t^d&=&\exp(\bm{x}_t\bm{\beta}),
\end{eqnarray}
where the row vector $\bm{x}_t$ contains the exogenous covariates at time $t$ and $\bm{\beta}$ is a column vector of coefficients. 
%Note in \eqref{eq:Poisson3} that $\bar{Y}$ is the maximum likelihood estimator for $E_t(\lambda_t)$, the expected value of the total rate of events. If $\lambda_t^d$ is expressed as a proportion of the expectation and the proportion is estimated using Bayesian methods, then \eqref{eq:Poisson3} is a semi-empirical Bayesian estimator of $\lambda_t^d$.  This improves the identifiability of the two components of the convolution model, improving the efficiency of the MCMC sampling scheme. 
%
%There are two terms in the summation in \eqref{eq:hawkes_intensity} where endogenous effects could be considered.  The decay function $g(\cdot)$ which has parameters determining the decay behaviour of past events' influence on the current rate is one option that is explored in \cite{Hamilton:Hamilton:1983}.\footnote{Contrary to the definitions posed in this paper \cite{Hamilton:Hamilton:1983} models the effects of factors such as GDP, societal openness, and education (exogenous factors) on the ``reversibility'' or decay parameter of the contagion process.}  The approach implemented here is to replace the term $Y_s$ in \eqref{eq:hawkes_intensity} with $Z_s$
%\mdp{Original:}The expected rate of the contagion process $\mu_t^c$ from \eqref{eq:gamma} is 
The expected rate of the contagion process $\mu_t^c$ from \eqref{eq:gamma} is 
\begin{eqnarray}
\mu_t^c&=&\sum_{s<t} \delta_s Y_s  g(t-s; \bm{\phi}).
\end{eqnarray}
The decay function $g(\cdot;\bm{\phi})$ controls how long (in days) the influence from past events persists, and is a non-negative function defined for the days $u\in\mathbb{N}$ such that $\sum_{1}^{\infty} g(u) = 1$. Following \cite{Porter:White:2012, White:Porter:2013}  $g(\cdot;\bm{\phi})$ is a shifted negative-binomial pmf, parametrised in terms of mean and scale.
The endogenous effects coefficient $\delta_s$ (also termed the \emph{volatility parameter}) is 
equal to the expected number of additional events generated by each event on day $s$.  As $\delta_s\geq 0$ the effects of covariates can be incorporated as in \eqref{eq:Poisson3} defining
\begin{eqnarray}
\label{eq:Zs}
\delta_s&=&\exp(\bm{w}_s\bm{\eta}),
\end{eqnarray}
where $\bm{w}_s$ is a row vector of endogenous covariates at time $s$ and $\bm{\eta}$ a column vector of coefficients; thus the contagion rate $\mu_t^c$ is the average of the number of events prior to time $t$ scaled by the endogenous effects coefficient and weighted by the decay function $g$. 

It is useful to note that the conditional expectation of $Y_t$
\begin{eqnarray}
\lambda_t|\lambda_t^d&=&\lambda_t^d+\lambda_t^c\\
&=&\lambda_t^d+\sum_{s<t} \delta_s Y_s  g(t-s; \bm{\phi})
\end{eqnarray}
is similar to the Hawkes self-exciting process intensity function \cite{Hawkes:1971a,hawkes71b}, but extends it by allowing $\lambda_t^d$ and $\delta_s$ to vary according to the exogenous and endogenous effects. 

\subsection{Selecting Endogenous and Exogenous Variables}
%Diffusion is defined as the changes in the rate of events due to precipitating events. 
%\mdp{ORIGINAL: 
%The covariates $\bm{x}_t$ for the endogenous effects can be defined based on known precipitating events or specified in order to allow a data-driven approximation, allowing the model to be used for exploratory purposes.  In this paper penalised $b$-splines \cite{Lang:Brezger:2001} are used for this purpose.}
%\gaw{(I accepted Mike's changes here)}\\
%\mdp{Proposed:}
Incorporating the exogenous effects into the diffusion term requires either an apriori knowledge of the precipitating events, or a flexible data-driven non-parametric model that can approximate their effects. 
The latter approach is followed in this paper to demonstrate the utility of this model as an exploratory tool for identifying possible precipitating events.  This is accomplished by using penalised $b$-splines \cite{Lang:Brezger:2001}, where the vector $\bm{x}_t$ in \eqref{eq:Poisson3} is a set of basis functions evaluated at time $t$.  
%
%The model in \cite{Hamilton:Hamilton:1983} only considers contagion as a mechanism for changes in rates of terrorism, but includes GNP, and indices of democracy or ``openness'' of societies, and measures of education as factors influencing the contagiousness of events (or in their terminology the ``reversibility'' of the rate increases).  The use of these variables is contrary to the definition of endogenous as applied to factors influencing the contagious process.  

%\mdp{Original:} Endogenous effects are based on characteristics of events that affect their ``contagiousness''.  Exploratory data analysis led to the selection of the number of fatalities as the variable describing the endogenous effects. The exact nature of the relationship between fatalities and contagiousness is not clear, therefore the covariates $\bm{w}_s$ in \eqref{eq:Zs} are defined as a penalised $b$-spline basis based on the natural log of the number of fatalities (plus 1), allowing a data-driven exploration of the relationship.   

%\mdp{Proposed:}
The endogenous covariates are constructed from the event characteristics that affect their ``contagiousness''.  Exploratory data analysis led to the selection of the number of fatalities as the variable describing the endogenous effects. Because the exact form of the relationship between fatalities and contagiousness is not known, the covariates $\bm{w}_i$ in \eqref{eq:Zs} are also modelled with penalised $b$-splines (over the natural log of the number of fatalities plus 1), allowing a data-driven exploration of the relationship.

%describes the endogenous effects in \eqref{eq:Zs}.  
%This use of penalised splines as in \eqref{eq:Poisson3} allows for the effect of fatalities to be non-linear. %a non-monotonic function. 

%Datasets 
%designed to capture this information such as the Correlates of War 
%(\texttt{http://www.correlatesofwar.org/}) , offer little guidance, and 
%collect data aggregated to a yearly level, limiting their usefulness in 
%modelling daily variation in rates.  Other sets such as \cite{QoG:2017} 
%(itself an aggregation of data from multiple sources) offer similar 
%resolutions, but include over 2000 possible covariates, introducing a 
%problem of variable selection or dimension reduction beyond the 
%scope of this paper.  In order to demonstrate the feasibility of the 
%approach suggested in \eqref{eq:Poisson3} an exploratory approach is 
%taken and $\bm{X}$ is defined as a Bayesian penalised spline 
%\cite{Lang:Brezger:2001} to identify trends in the diffusion process 
%data.

%  For the contagion process the list of possible endogenous covariates is much smaller.  

\section{Computation and Implementation\label{comp}}
%% Computation and implementation
%The data are observed as $Y_t$, the total number of events in each interval $t$ (a single day).  The total number of events are assumed to be the sum of events from the diffusion and the contagion processes, $Y_t=Y_t^d+Y_t^c$.  
As discussed in Section \ref{Model} 
%because $Y_t^d$ is assumed to follow a Poisson distribution and $Y_t^c$ is assumed to follow a negative-binomial distribution, there is no closed form for the convolution of these two processes, thus no closed form for the likelihood of $Y_t$.   This makes evaluating the model in either a frequentist or Bayesian context computationally challenging.  If the partition of the data $Y_t$ into $Y_t^d$ and $Y_t^c$ is known then the individual likelihoods for for the diffusion and contagion processes can be used to estimate their respective parameters.  
the data can be partitioned stochastically using \eqref{eq:binom} at each iteration of an MCMC scheme to sample from the posterior distribution of the model parameters. The MCMC scheme can be completed given prior distributions for the parameters $\bm{\phi},\bm{\beta},\bm{\eta}$ and $\sigma^2$.  
%The model in \eqref{eq:Poisson} -- \eqref{eq:binom} and  \eqref{eq:mu_c} -- \eqref{eq:lambda_d} describes the likelihood and the forms of the rates for the diffusion and contagion processes.  The details of evaluating this model to estimate its parameters are as follows.  
%\subsection{The Likelihood}
%The likelihood of the observed data is necessary in order to estimate model parameters and make inference on their estimates.  In a frequentist setting it is used directly, and in the Bayesian context demonstrated here, it is a necessary component of the posterior density.  
%  by noting that while the marginal distribution of $Y_y^c$ is negative binomial, is can be shown that if the mean of the negative-binomial follows a gamma distribution then the conditional distribution of $Y_t^c$ can be shown to be a Poisson distribution, resulting in the conditional density for $Y_t$ being a Poisson distribution.  As a result given the diffusion and contagion process rates 
\subsection{Priors}
The likelihood for the partitioned data are derived from \eqref{eq:Poisson2} and \eqref{eq:nb}, and the Bayesian model is completed by specifying prior distributions for the parameters.  The parameters $\bm{\beta}$ and $\bm{\eta}$ are given penalised first-order random walk priors as defined in \cite{Lang:Brezger:2001} which encourages parsimony and discourages over-fitting by penalising the differences between parameter values, i.e. for a set of spline basis functions $\bm{X}$ if all the parameter values $\bm{\beta}$ are equal, then $\bm{X\beta}$ is a constant. This expresses the prior belief that the diffusion rate is constant over time and that there is no relationship between the number of fatalities and the contagiousness of an event. 
\begin{eqnarray}
\pi(\bm{\beta}|\rho)&\propto&\exp\left(-\frac{\rho}{2}\bm{\beta}'\bm{K}\bm{\beta}\right)\\
\pi(\bm{\eta}|\gamma)&\propto&\exp\left(-\frac{\gamma}{2}\bm{\eta}'\bm{B}\bm{\eta}\right)
\end{eqnarray}
where $\bm{K}$ and $\bm{B}$ are first-order random walk penalty matrices, e.g.  
\begin{eqnarray*}
\bm{K}&=&\left(\begin{array}{lllll}
\phantom{-}1 & -1 & & & \\
 -1 & \phantom{-}2 & -1 & & \\
& \ddots & \ddots & \ddots &  \\
 & & -1 & \phantom{-}2 & -1\\
 & & & -1 & \phantom{-}1
\end{array}\right).
\end{eqnarray*}
%These matrices are rank deficient, thus improper, but  the resulting posteriors are proper.  
Vague proper gamma distributions are specified as conjugate hyper-priors for $\rho$ and $\gamma$ as suggested in \cite{Lang:Brezger:2001}.  
%\begin{eqnarray}
%\rho&\sim&\\
%\gamma&\sim&
%\end{eqnarray}
%These are conjugate priors resulting in gamma conditional posterior densities.  
The remaining priors are
\begin{eqnarray}
\pi(\bm{\beta})&\propto&1\\
\pi(\bm{\eta})&\propto& 1\\
\pi(\sigma^2)&=&\frac{1}{(1+\sigma^2)^2}\\
\pi(\bm{\phi})&=&\Pi_{i=1}^2\frac{1}{(1+\phi_i)^2}.
\end{eqnarray}

\subsection{Computation}
%The model is evaluated using an MCMC scheme based on a Gibbs' sampling scheme draw samples from the resulting posterior distribution via the conditional posterior distributions making up the full conditional.  
Mixture models can be difficult to evaluate as identifiability issues can cause poor mixing or convergence of MCMC chains \cite{Robert:Casella:2004}. 
%The re-parametrisation of the model using \eqref{eq:binom} to partition the data, and the parametrisation of $\lambda_t^d$ in \eqref{eq:lambda_d} help with the mixture and convergence issues, in addition there are 
This is addressed using additional re-sampling steps for $\bm{\beta}$ and $\bm{\eta}$ as suggested in \cite{Rouder:etal:2003}. For example, for $\bm{\beta}$ at the $i$th iteration:
\begin{itemize}
\item[i)] Sample $\bm{\beta}^{(i)}\sim\pi(\bm{\beta}|\rho, \bm{Y^d})$
\item[ii)] Sample $\bm{\beta}^*\sim N(\bm{\beta}^{(i)},\alpha \bm{\mathcal{I}})$
\item[iii)] Let $\bm{\beta}^{(i)} = \bm{\beta}^*$ w.p. $\min\left(1,\frac{\pi({\bm{Y^d}|\bm{\beta}^*})}{\pi(\bm{Y^d}|\bm{\beta}^{(i)})}\right)$.
%\item[iv)] Else let $\bm{\beta}^{(i)}=\bm{\beta}^{(i)}$.
\end{itemize}
The value $\alpha$ is chosen to set the acceptance rate to between 20\% -- 40\%, as recommended in \cite{Rouder:etal:2003}.
The rest of the implementation is a straightforward Gibbs MCMC scheme using a hit-and-run sampler with slice sampling \cite{Smith:1996} for those parameters without conjugate conditional posterior distributions.   

\section{Results}
\begin{figure}[htbp]
 \centering
\includegraphics[height=0.33\textheight]{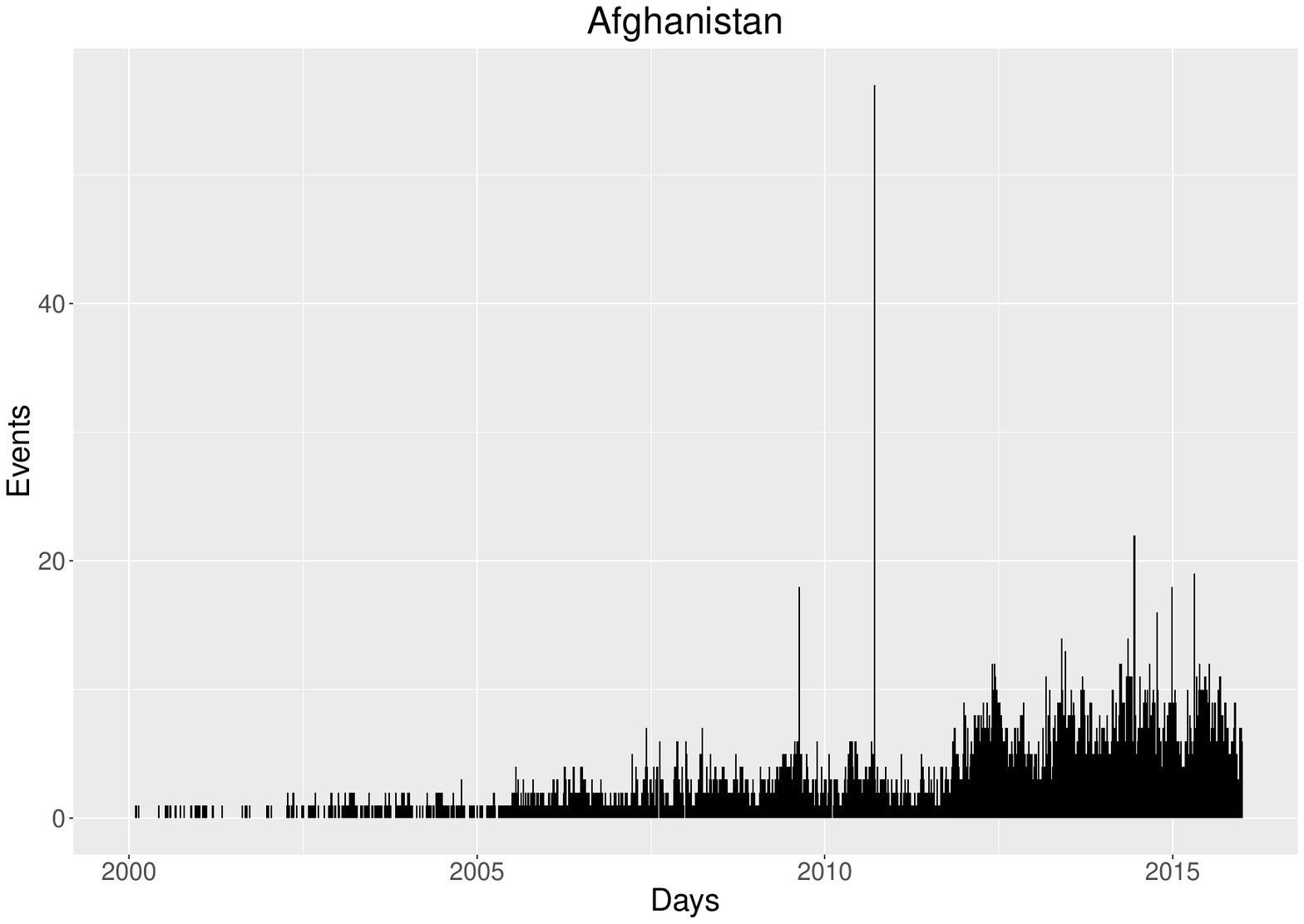}
\caption{\label{afghan_data}Events in Afghanistan 2000-2016}
\end{figure}

\begin{figure}[htbp]
 \centering
\includegraphics[height=0.33\textheight]{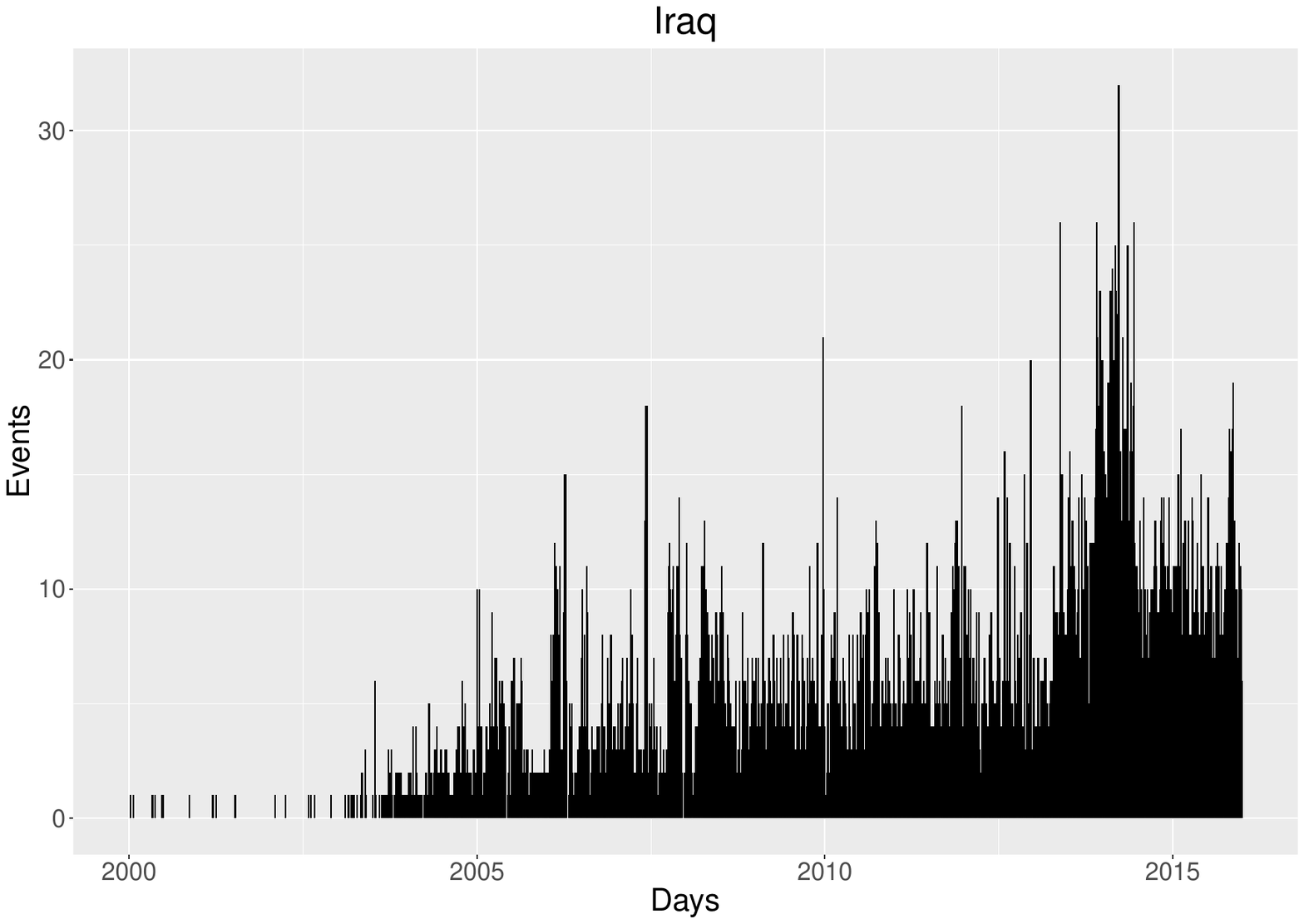}
\caption{\label{iraq_data}Events in Iraq 2000-2016}
\end{figure}

\begin{figure}[htbp]
 \centering
\includegraphics[height=0.33\textheight]{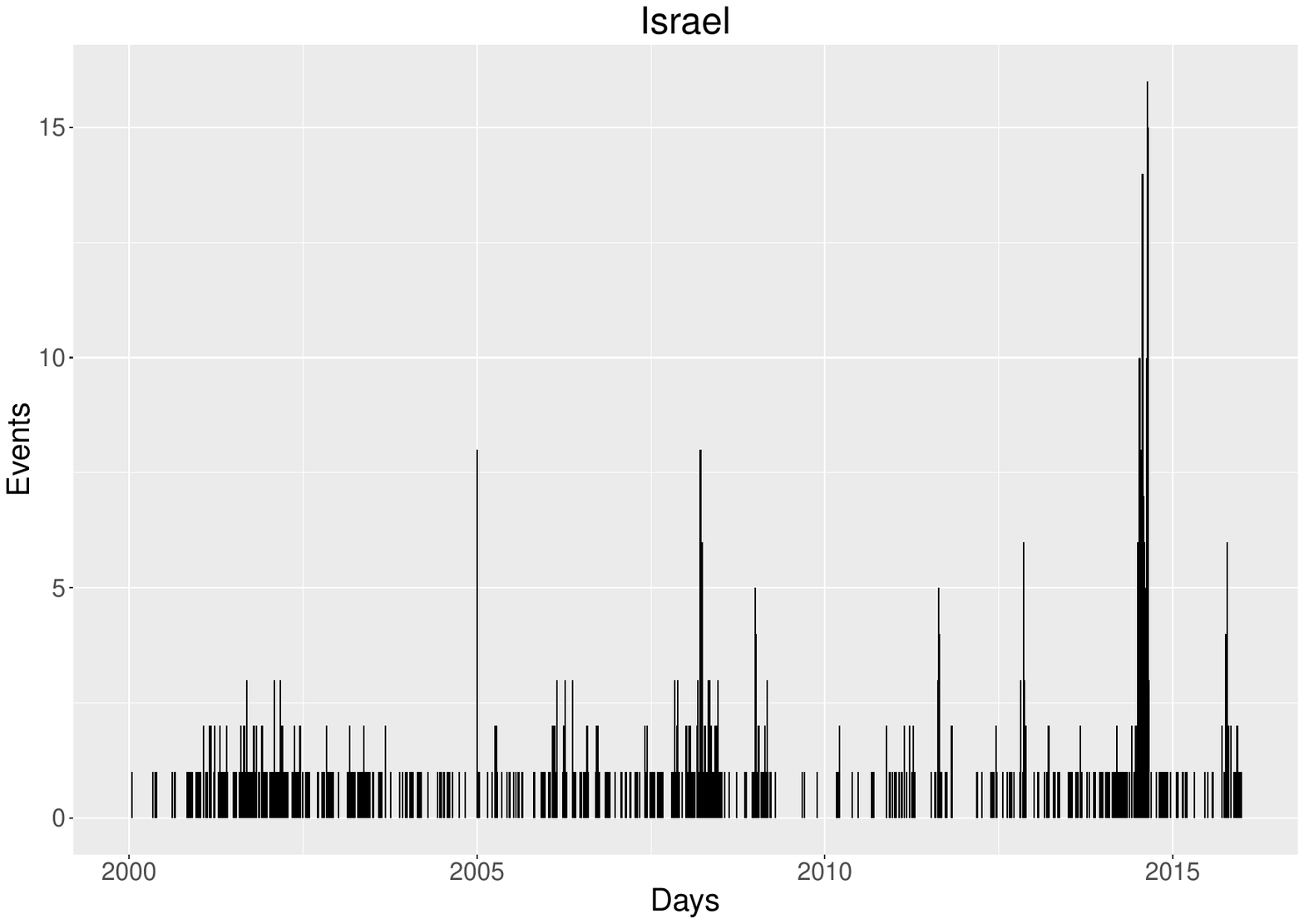}
\caption{\label{israel_data}Events in Israel 2000-2016}
\end{figure}

The Global Terrorism Database (GTD) is an open-source, publicly available dataset that contains records of terrorist events from 1970 \cite{GTD:2017}, containing as at its most recent reporting year, 2016, over 170,000 cases each with up to 120 variables.  
%The GTD is currently the most comprehensive unclassified database of terrorist events in the world 
GTD data from the period 2000--2016 from the countries of Afghanistan, Iraq, and Israel, shown in Figures \ref{afghan_data}, \ref{iraq_data}, and \ref{israel_data} were analysed using the model derived in Sections \ref{Model} and \ref{comp}.  
%
%
%Comprehensive analysis of these data requires an in-depth understanding of historical context and geo-politics; naturally the analyses here are limited to results and inferences that can be drawn from the model results.  
%Where possible references are made to both the recent historical context of the data and key events or points in time in order to augment the model analyses. 
Results indicate that the diffusion and contagion processes are identifiable and demonstrate distinct patterns that align with theoretical constructs and the historical narrative. Furthermore, the model suggests that the excitation effect varies significantly with the number of fatalities.  

%\textcolor{red}{- show counts plot and/or data summary (e.g. \# events, ybar, etc)}

\subsection{Afghanistan}
\begin{figure}[htbp]
 \centering
\includegraphics[height=0.33\textheight]{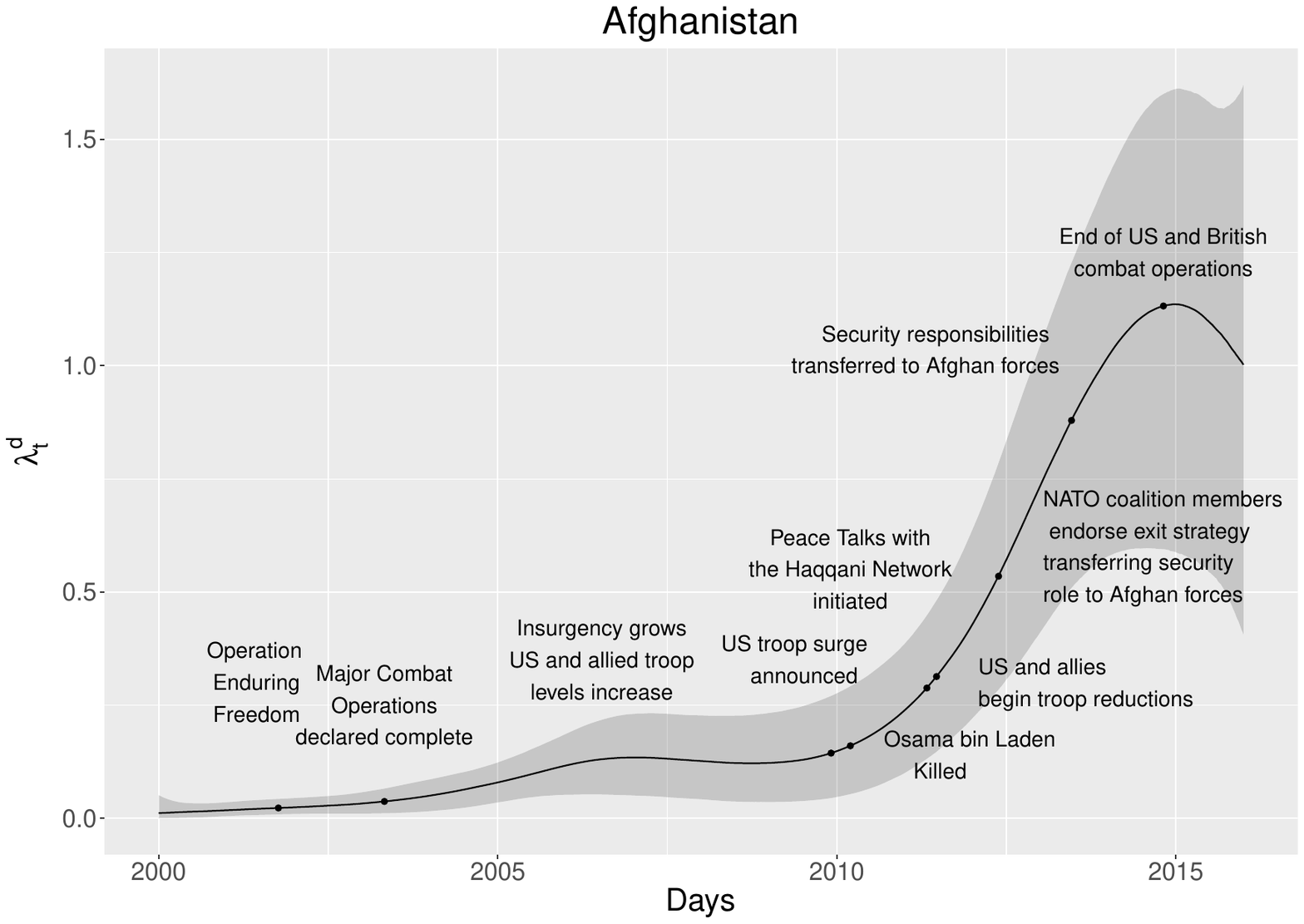}
\caption{\label{afghan_mu}$\mu_t^d$ for Afghanistan 2000-2016}
\end{figure}

%According to the definition in \cite{Midlarsky:1978} changes in the diffusion rate are linked specific events or precipitants, as suggested in \cite{Crenshaw:1981}.   
%The diffusion rate is proportional to $\alpha_t$ making it a useful variable to explore the diffusion process, especially in comparison to $p_t^d$ as shown in Figure \ref{afghan_pcb}. 
Figure \ref{afghan_mu} shows the median and 95\% credible interval of the diffusion rate $\mu_t^d$ for Afghanistan from 2000 through 2016, with some key events annotated. 

The diffusion rate is quite low in the period of de facto Taliban rule over a majority of Afghanistan from 1998 until the initial US-led invasion in October 2001 \cite{Rashid:2000}.
After the invasion, activity begins to slowly increase as the Taliban began re-organising and re-grouping, and beginning an insurgency and engaging in guerilla warfare \cite{Tohid:Scott:2003,Carlotta:2004}.
By 2006 US forces were replaced by NATO coalition forces in southern Afghanistan, with the goal of forming Provincial Reconstructions Teams to begin rebuilding Afghanistan and stabilise the political situation \cite{Bebber:2008}.  
Multiple operations by US and NATO forces to push Taliban forces out of the provinces met with varying degrees of success over the next few years as the US increased troop levels by over 80\% in an attempt to defeat the Taliban \cite{OBryant:Waterhouse:2008}.  
Despite these efforts, by the end of 2009 the Taliban's strength had returned to near pre-invasion levels \cite{OHanlon:2010} and intelligence showed a steady increase in security incidents \cite{Bergen:2010}.  
A slight decrease in the rate of incidents in 2010 coincides with the initiation of peace talks with the Haqqani network by Hamid Kharzai in March of 2010 and the Afghan Peace Jirga \cite{Reuters:2010}.  
Increases in US troop levels continued in 2010 as part of a ``surge''  strategy with shift to target Taliban leadership resulting in the capture or killing of more than 900 low- to mid- level Taliban leaders \cite{VandenBrook:2011}.  
There is a sharp upturn in $\mu_t^d$ starting in 2011 and coinciding with the death of Osama Bin Laden in May 2011 and the announcement of US troop withdrawals \cite{Landler:Cooper:2011}, followed by similar announcements and withdrawals by other coalition members. 
As the diffusion rate continued to increase, in May of 2012 NATO coalition members endorsed an exit strategy transferring responsibility for security to Afghan forces by mid-2013 \cite{Reuters:2012,Felbab-Brown:2012} which occurred on 18 June 2013 \cite{Hodge:2013}.  
On 26 October 2014 Britain and the US formally ended their combat operations, handing over their last remaining bases to Afghan control \cite{Syal:2014,Johnson:etal:2014}.

%The fitted model also assumed that the degree of contagiousness of events was a function of the number of fatalities.

\begin{figure}[htbp]
\centering
\includegraphics[height=0.33\textheight]{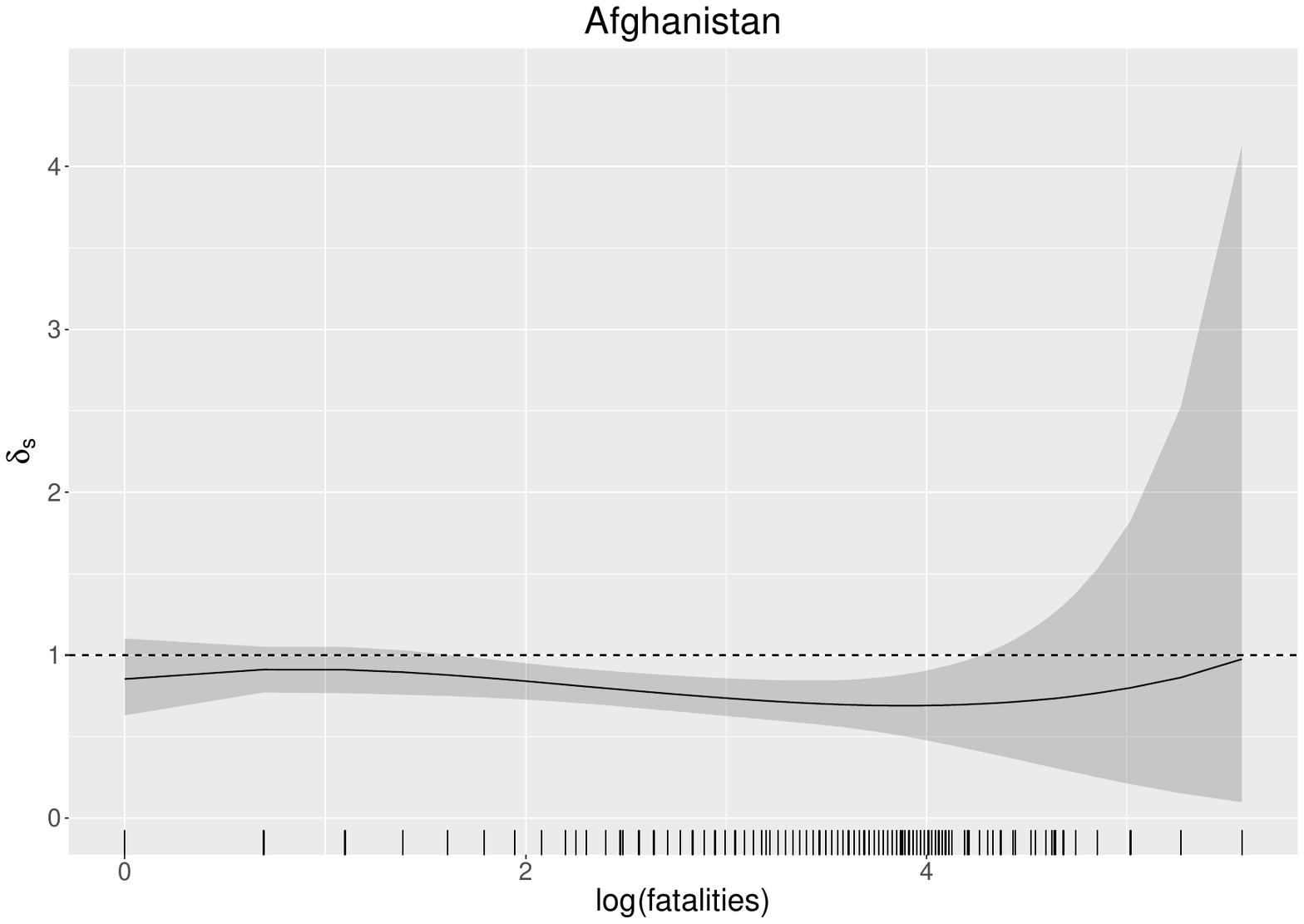}
\caption{\label{afghan_nkill}$\delta_s$ vs. Fatalities in Afghanistan 2000-2016}
\end{figure} 
The contagion process is best explored in detail through the concepts of volatility and resilience as defined in \cite{White:etal:2012b}. Volatility is the expected number of events that occur via contagion after each event, this is parametrised as $\delta_s$. Figure \ref{afghan_nkill} shows the median and 95\% credible interval for the endogenous effects coefficient $\delta_s$ as a function of the number of fatalities due to events at time $s$. 
When the expected volatility is greater than $1$, each event is expected to produce more than one subsequent event and the contagion process becomes non-stationary or ``explosive". 
%In this case the contagion process is non-stationary or ``explosive''.
%of excitation effect of events relative to the number of fatalities expressed as the ratio $Z_s/Y_s$. 
The values for the median of $\delta_s$ provide evidence of a stationary contagion process (i.e. a volatility consistently less than $1$). There is an upward trend in the excitation effect for events with greater than 50 fatalities, but the limited amount of data reflected in the large credible intervals at the limits, makes it difficult to verify.  
Considering the number of fatalities as a measure of attack size, and an indirect measure of the resources allocated towards an attack, $\delta_s$ becomes a measure of operational capability.  Assuming that large attacks are a substantial drain on resources to mount future attacks, then $\delta_s$ would be expected to decrease with the number of fatalities.  The fact that it is approximately constant would seem to indicate that large attacks are not a significant drain on resources or capacity. The increase in frequency \emph{and} lethality of attacks over time, indicated by the near constant values of $\delta_s$, reflects the Taliban's resurgence during this time period.  
%is a reflection of the increase over time of the Taliban's relative strength.  
%This is likely a reflection that the frequency and lethality of events tends to increase over time, reflecting an increase in relative strength of the Taliban (and other insurgent groups) resurgence.  
%This also explains the observed increase in the excitation effect for higher fatality events, though the overall rarity of high fatality events makes this difficult to assess as there are only ten events with more than 100 fatalities, as evidenced in the increasing width of the credible interval.  
%
\begin{figure}[htbp]
\centering
\includegraphics[height=0.33\textheight]{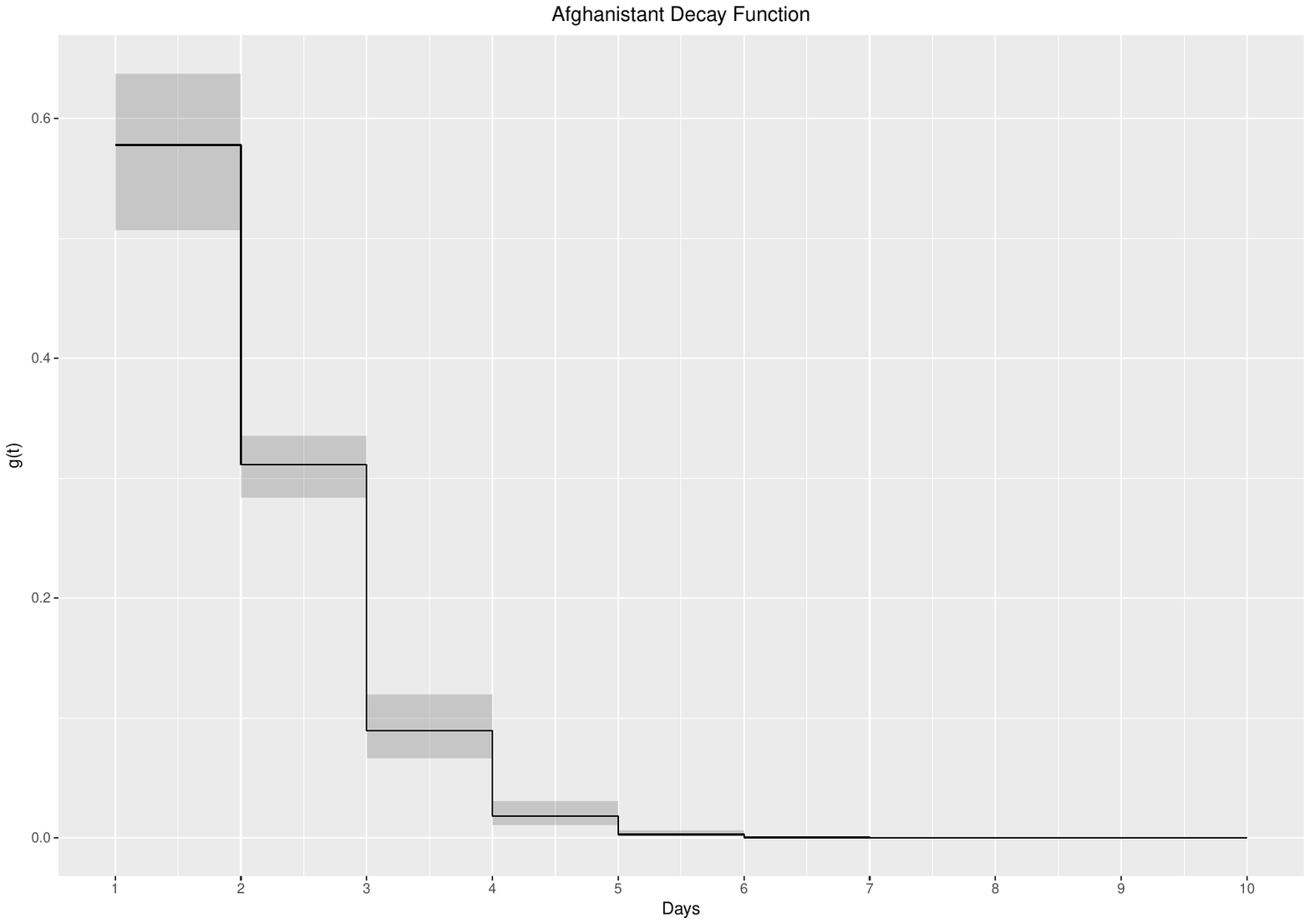}
\caption{\label{afghan_decay} Estimated decay function, $g(\cdot;\bm{\phi})$, for Afghanistan 2000-2016.}
\end{figure} 
 The parameters of the decay function $g(\cdot)$ (see \ref{IncEffects}) describe resilience, or the duration and intensity of the contagion effect.  
 The contagion effect can be explained using the language of \cite{hawkes&oakes:1974} describing the self-exciting model as a cluster process where events serve as a ``parent'' producing ``children'' through the contagion process (which subsequently become ``parents'' have ``children'' of their own). 
 The decay function for Afghanistan in Figure~\ref{afghan_decay} (median with 95\% credible interval) shows that the duration of the contagion effect of an event is limited to a few days. The expected time until a ``child'' event (i.e. an event attributed to the contagion effect of a previous, or ``parent'' event) is between $1.46$ and $1.70$ days after the ``parent'' event ($95\%$ credible interval) with an expected value of $1.6$ days. The probability of a ``child'' event occurring more than $3$ days after the originating event is less than $0.05$.

\subsection{Iraq}

\begin{figure}[htbp]
\centering
\includegraphics[height=0.33\textheight]{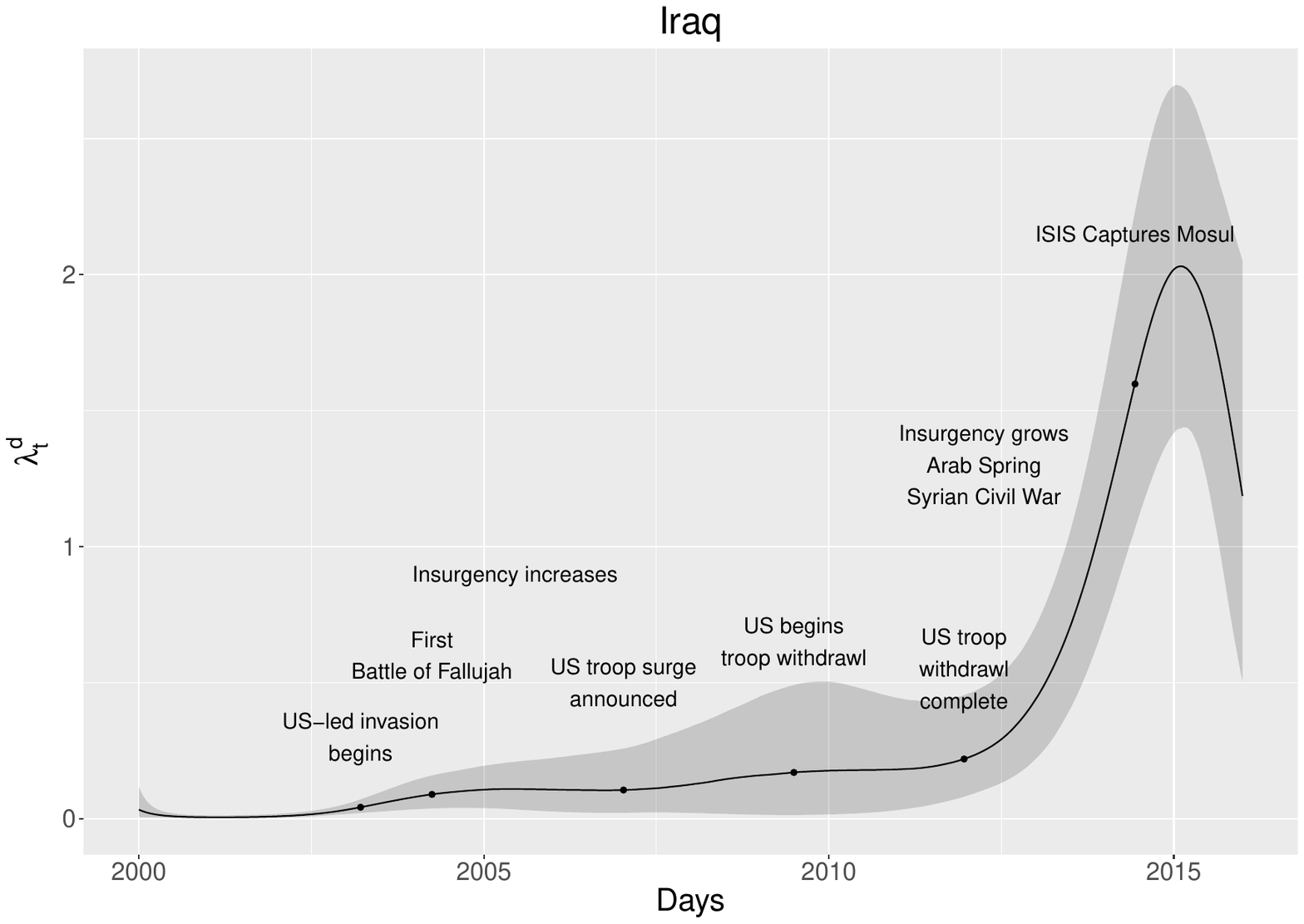}
\caption{\label{iraq_mu}$\lambda_t^d$ for Iraq 2000-2016}
\end{figure}

Figure \ref{iraq_mu} shows the posterior median and 95\% credible interval for the diffusion rate $\mu_t^d$ in Iraq for the period 2000--2016.  

Similar to Afghanistan there were few events in Iraq prior to the US-led invasion in 2003.  
%This is not surprising considering the ruling governments lack of transparency in reporting, but it is also suspicious given the long running conflicts with ethnic Kurds in northern Iraq and the regime of Saddam Hussein. 
The US-led invasion of Iraq began on 20 March 2003 and proceeded rapidly with the capital of Baghdad falling to US troops on 9 April 2003 and the declaration of the end of major combat operations on 1 May 2003 \cite{Keegan:2005}.  
Saddam Hussein remained at large until 13 December 2003, and significant pockets of resistance remained despite the coalition's efforts to establish a stable post-invasion democracy \cite{Hashim:2005}. 
After the end of conventional fighting, an insurgency began. Initially fuelled by Ba'ath Party loyalists the insurgency soon drew religious radicals and regular citizens.  
The violence came to a head on 31 March 2004 when insurgents in Fallujah captured and killed four US private military contractors resulting in the First Battle of Fallujah from 1 April 2004 until 1 May 2004, a campaign to secure Fallujah and capture the insurgents responsible for the deaths of the US contractors \cite{CNN:2004}. 
The Second Battle of Fallujah from 7 November 2004 to 23 December 2004 followed as some of the bloodiest fighting in Iraq, and the first time US forces fought forces made up exclusively of insurgents and not the remnants of the Iraqi Republican Guard \cite{Ricks:2007}. 
In 2005 the Iraqi Transitional Government was elected and charged with writing a new constitution.  
Despite this insurgent activity continued to escalate, including attacks on the Abu Ghraib prison and fighting around Baghdad and the Euphrates valley \cite{Ricks:2007}.   
In January of 2007 President Bush announced an increase in both troop levels and reconstruction efforts, this ``surge'' strategy under the command of the newly appointed commander of the Iraq Multinational Force, Gen. David Petraeus \cite{Bush:2007}.  
In March of 2007 the Iraqi Parliament enacted legislation calling on the US to set a timetable for withdrawal of their forces \cite{AP:2007}.  
By September of 2007 plans were in place to reduce US troop levels to pre-surge numbers \cite{Flaherty:2007}. 
On 4 December 2008, the Iraqi government approved the US-Iraqi Status of Forces Agreement requiring US forces withdraw from Iraqi cities by 30 June 2009 and all US forces be out of Iraq by 31 December 2011.  
US forces began their withdraw at the end of June 2009, handing over 38 bases to Iraqi control and removing all forces from Baghdad.  
In October of 2011 the departure of the remaining US troops was announced, and on 18 December 2011 the last US troops left Iraq.  
After the withdrawal of US troops the insurgency increased dramatically as Sunni militant groups stepped up attacks of the Shia majority \cite{Knights:2011}. 
Between 2011 and 2013 the Arab Spring inspires uprisings across the Arab world, including Syria where it ignites a civil war and gives rise to the Islamic State \cite{Kareem:Hwaida:2013,Ruthven:2016}.

\begin{figure}[htbp]
\centering
\includegraphics[height=0.33\textheight]{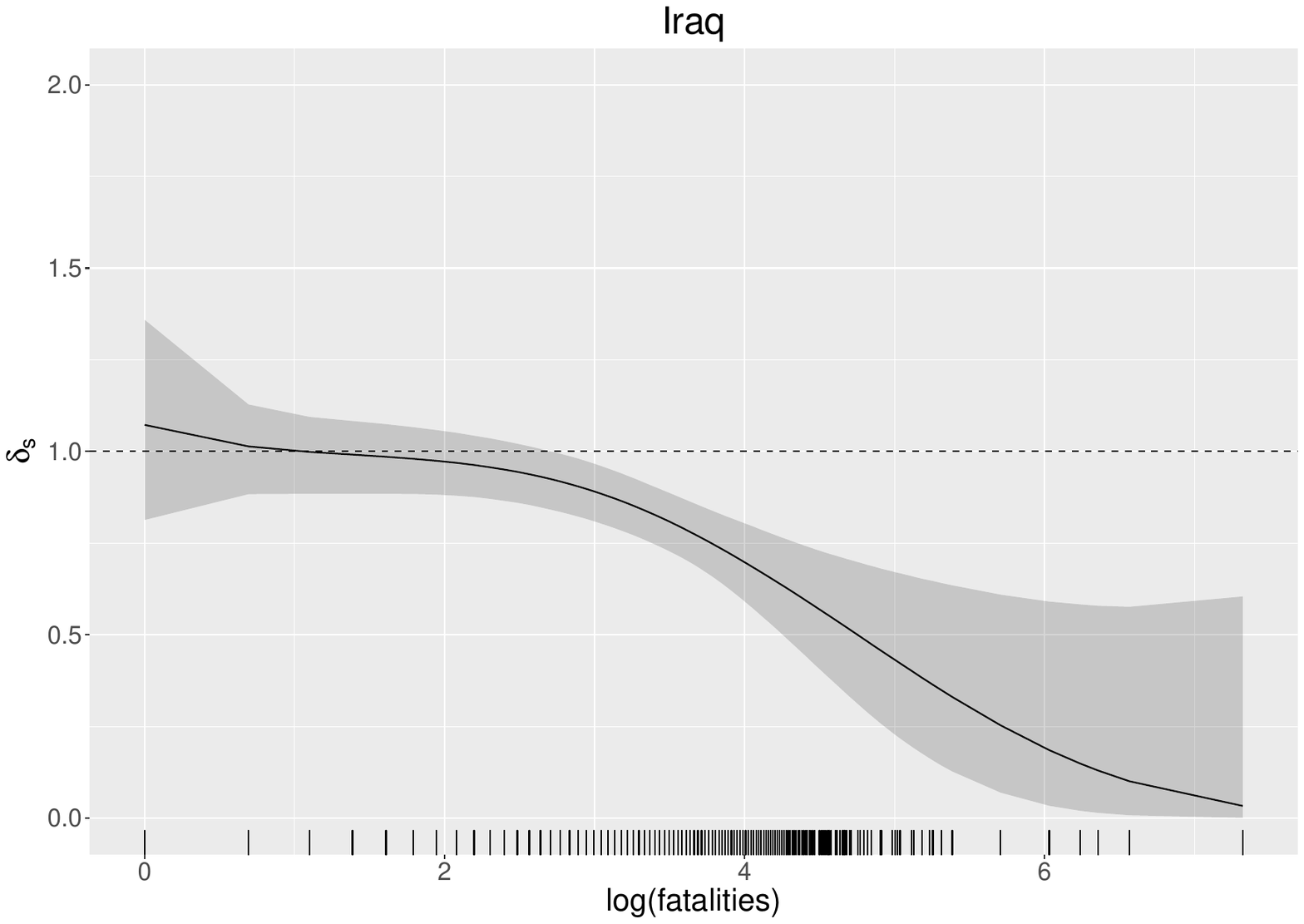}
\caption{\label{iraq_nkill}$\delta_s$ vs. Fatalities in Iraq 2000-2016}
\end{figure} 

The variation in volatility as a function of fatalities, as shown in Figure~\ref{iraq_nkill}, is pronounced.  
There is an obvious trend indicating that volatility decreases as the number of fatalities increases. The model suggests that events producing a large number of fatalities tend to generate almost no contagion effects. 
However, for events causing $0$ or $1$ fatalities the median value of $\delta_s$ is greater than $1$ indicating that the contagion is potentially in a non-stationary, explosive state. 
The trend for the frequency of events and the number of fatalities is not as strong as it is in Afghanistan, and there are  38 events with over 100 fatalities, leading to a narrower credible interval at the extremes for $\delta_s$.  This paints a different picture of the operational capacity.  Rather than the steady increase as seen in Afghanistan by the resurgence of the Taliban, a power vacuum was created after the invasion.  The coalition forces outlawed the ethnic minority Ba'ath party and the disbanded the Iraqi armed forces, many of whom were eventually integrated into the Islamic State\cite{Shadid:2010}. Initially, the resistance to the occupation had no central unifying entity, and it wasn't until the rise of the Islamic State post-2010 and their subsequent assimilation of other resistance groups that a similar unified force against the occupation existed\cite{MERMI:2013}. This explains both the decrease in volatility with casualties, and the non-stationarity for events with $0$ or $1$ fatality.  
%where each event is expected to generate more than one additional event.  
%There is significant decrease in the excitation effect for events as the number of fatalities increase.  The credible interval doesn't show the same increase in width as in Afghanistan. 
%While there is a similar, but smaller correlation between the frequency of attacks and the number of fatalities, there are more large fatality events (38 events record over 100 fatalities).  
%
\begin{figure}[htbp]
\centering
\includegraphics[height=0.33\textheight]{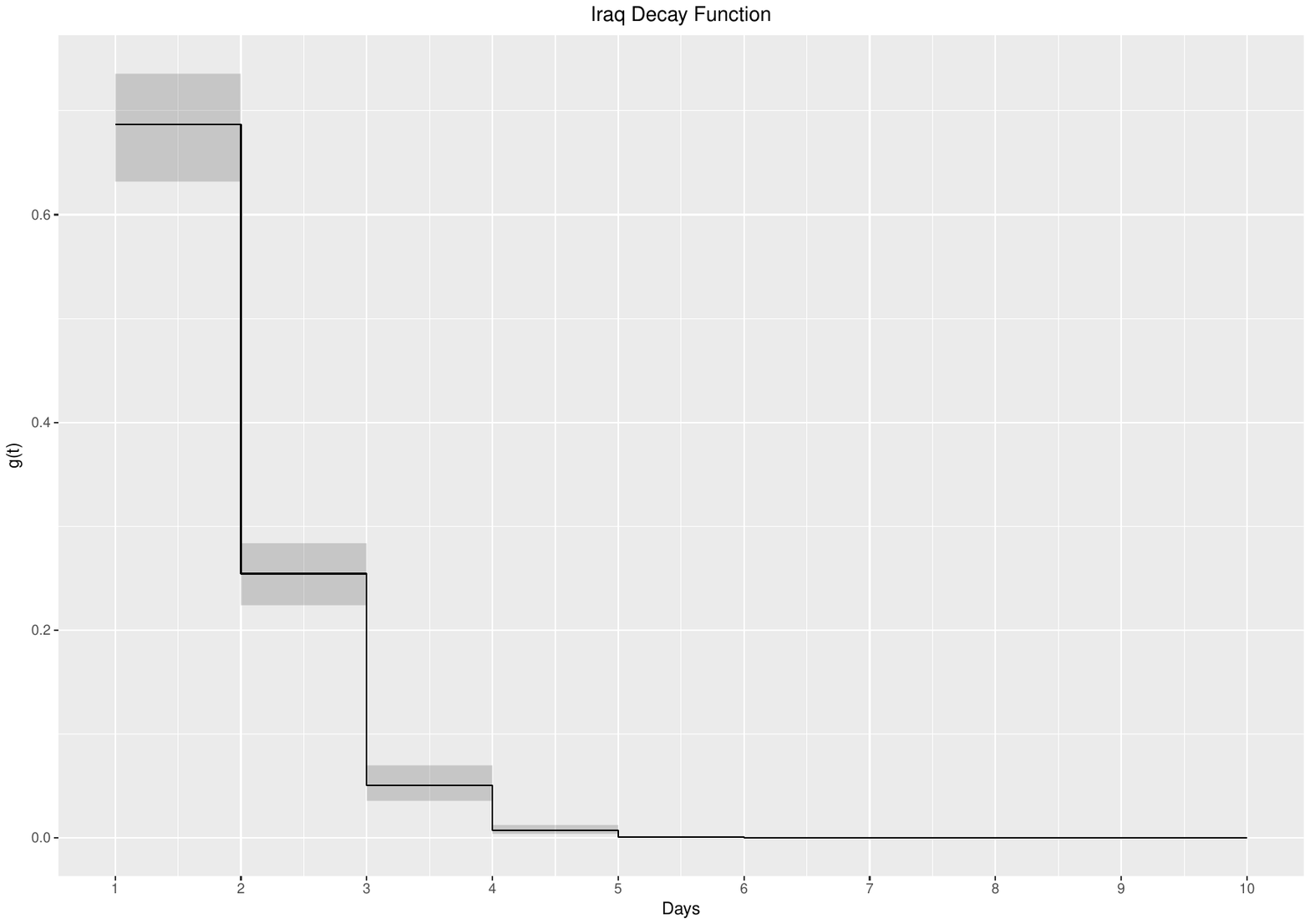}
\caption{\label{iraq_decay} Estimated decay function, $g(\cdot;\bm{\phi})$, for Iraq 2000-2016.}
\end{figure} 
The resilience of activity in Iraq is described in the decay function $g(\cdot)$ shown in Figure~\ref{iraq_decay} (median with 95\% credible interval) and the parameters of the function $g(\cdot)$.  The duration of contagion is limited to a few days, the expected time until a contagion event is $1.4~(1.31,1.47)$ days and the probability of a contagion event occurring more than $3$ days after the originating event is less that $0.05$. 

%\textcolor{red}{- show decay function; report mean/median etc. Shift by one day. }

\subsection{Israel}
\begin{figure}[htbp]
\centering
\includegraphics[height=0.33\textheight]{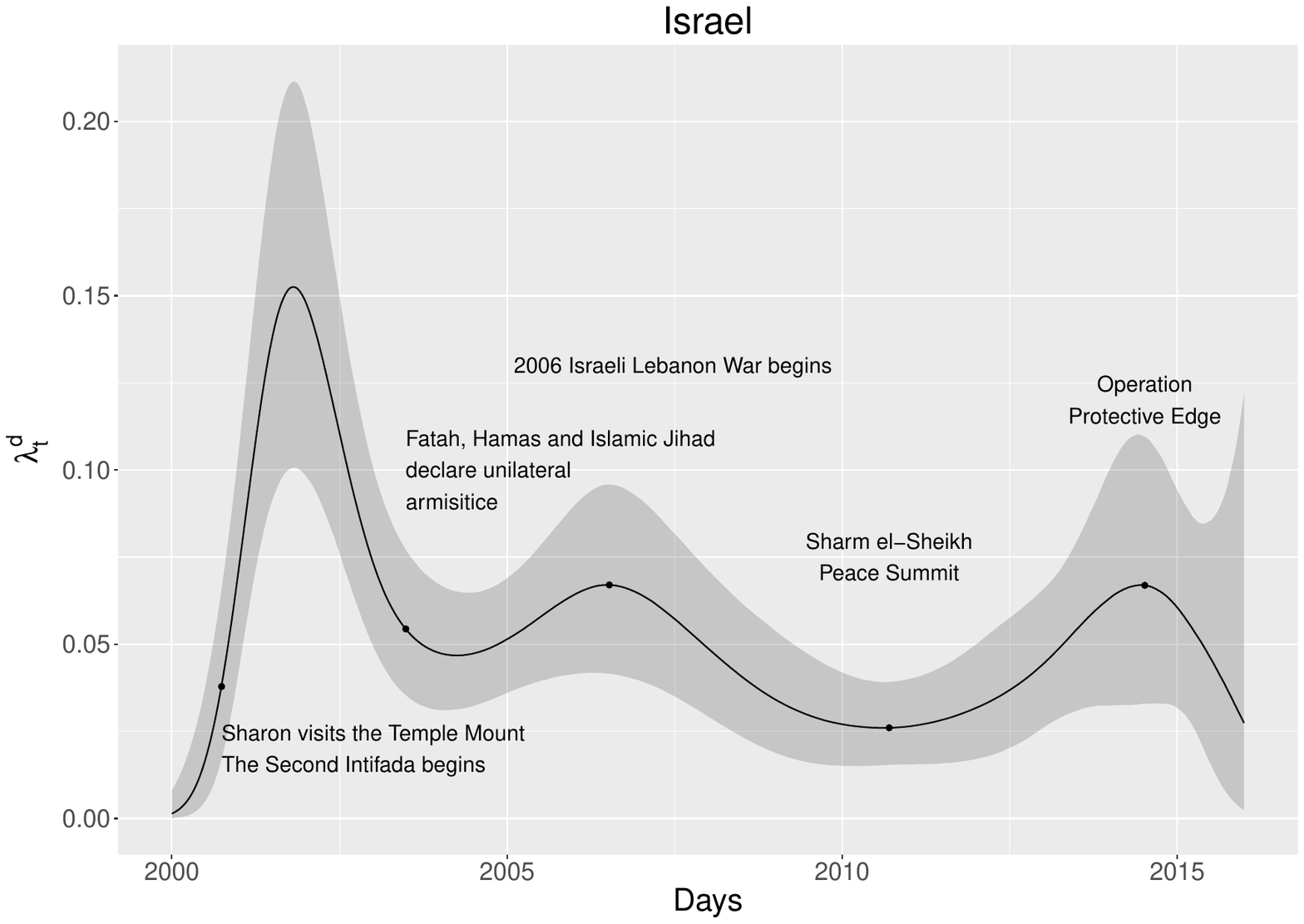}
\caption{\label{israel_mu}$\lambda_t^d$ for Israel 2000-2016}
\end{figure}

The history, context, and model results for Israel differ from those of Afghanistan and Iraq. This is reflected in Figure \ref{israel_mu} showing the posterior median and 95\% credible interval of $\mu_t^d$ for Israel from 2000 to 2016.  
There is significant variation in the diffusion rate, but it never reaches the intensity of Afghanistan and Iraq. 
There are numerous events that could be identified as possible precipitants, a few have been noted here that align with the data to provide some reference to the results. 
There is a steep increase in $\mu_t^d$ following Prime Minister Sharon's visit to the Temple Mount in Jerusalem \cite{Goldberg:2000}.
Shortly after this visit the Second Intifada began, a period of increased Israeli-Palestinian violence, and the peak value of $\mu_t^d$, of approximately $0.15$, is reached.  
After this peak there is a steady decrease surrounding a three-month cease fire announced by Fatah, Hamas and Islamic Jihad on 29 June 2003 \cite{Bennet:2003}.
There is a second peak in $\mu_t^d$ near the start of the 2006 Israel-Lebanon War \cite{Myre:Erlanger:2006}.  
The lowest value for $\mu_t^d$ is reached in September of 2010 around the time of the Sharm el-Sheikh peace summit \cite{Haaretz:2010}.
After 2010 there is again an increase in $\mu_t^d$ culminating around the time Israel launched Operation Protective Edge, to stop the launch of rockets into Israel from Gaza \cite{Thrall:2014}.

% The variation in $\alpha_t$, particularly the large increase around 2001, is associated with the beginning of the second(?) intifada in late 2000, in the increase end with the declaration of a unilateral armistice by Fatah, Hamas, and Islamic Jihad in late-June of 2003.  A smaller increase follows the kidnapping of an Israeli solder in in June 2006, subsiding with an Israeli-Hamas ceasefire in 2008.  Another increase coincides with Operation Protective Edge (the 2014 Israel-Gaza conflict) launched in early July 2014 by Israel after the kidnapping and murder of three Israeli teenagers by members of Hamas\cite{Khoury:2014}. 

\begin{figure}[htbp]
\centering
\includegraphics[height=0.33\textheight]{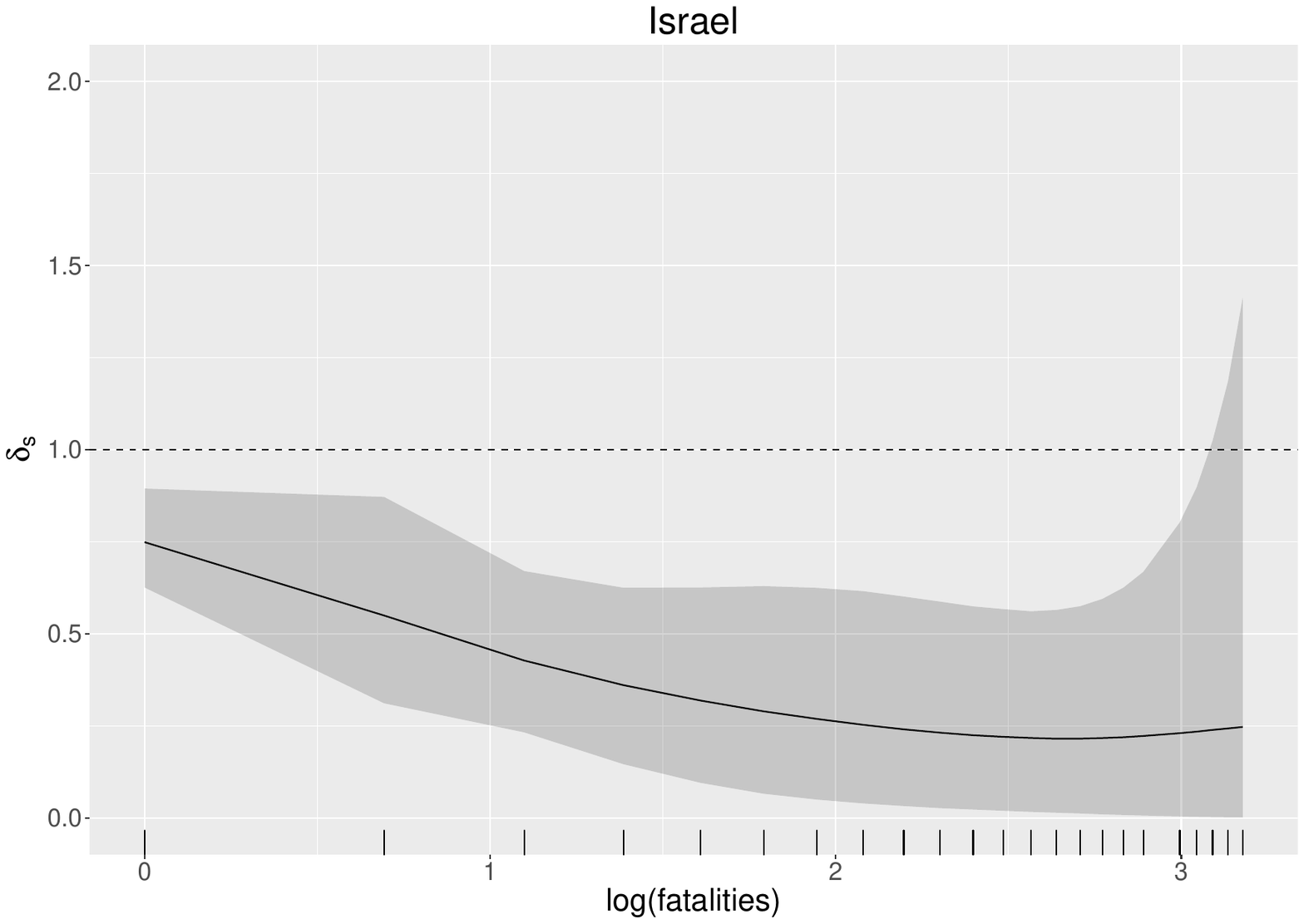}
\caption{\label{israel_nkill}$\delta_s$ vs. Fatalities in Israel 2000-2016}
\end{figure} 

The variation of the volatility as measured by $\delta_s$ for events in Israel follows a similar  decreasing pattern as in Iraq and Afghanistan.
%, it is decreasing in the number of fatalities across the range.  
The volatility never exceeds $1$ indicating that the contagion process is stationary. 
%There is a similar widening of the credible interval at the highest level of fatalities, but there is no increase in the median value.  
While the number of fatalities per event tends to increase over time in Afghanistan, it is decreasing in Israel over the same period.  
\begin{figure}[htbp]
\centering
\includegraphics[height=0.33\textheight]{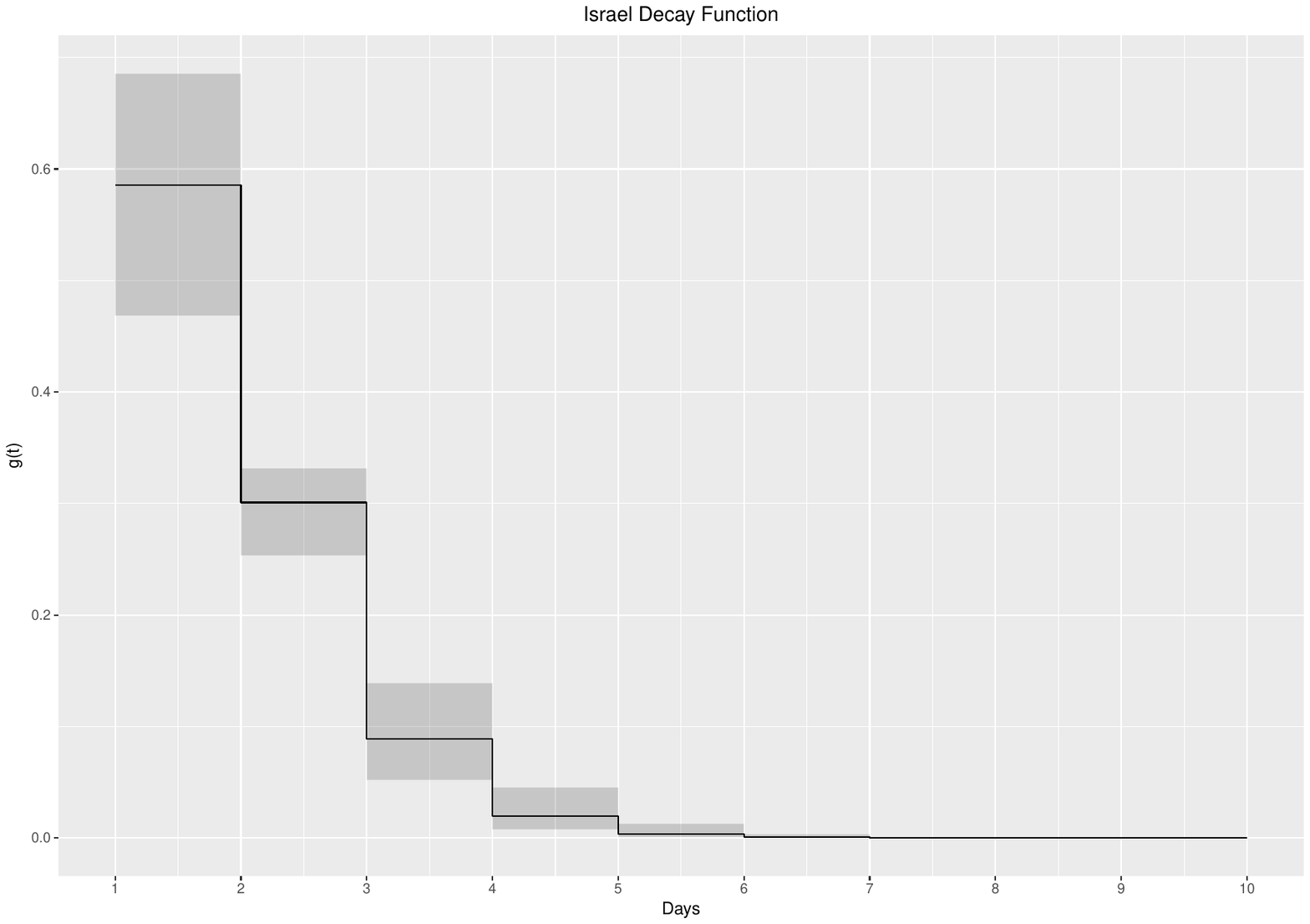}
\caption{\label{israel_decay} Estimated decay function, $g(\cdot;\bm{\phi})$, for Israel 2000-2016.}
\end{figure}
The decay function in Figure~\ref{israel_decay} (median with 95\% credible interval)  shows that the duration of contagion in Israel is limited to a few days.  Specifically, the expected time until a contagion event is around $1.6~(1.39,1.82)$ days and the probability of a contagion event occurring more than 4 days after the originating event is less than $0.05$.

%\textcolor{red}{- show decay function; report mean/median etc. Shift by one day. }

%\subsection{Diffusion Effects}
%The diffusion rate $\lambda_y^d$ is expressed in \eqref{eq:lambda_d} as a proportion of $\bar{Y}$, this means that variation over time in the rate of diffusion events can be examined in both absolute terms ($\lambda_t^d$) or in relative terms ($\alpha_t$).  Posterior estimates of $\lambda_t^d$ and $\alpha_t$ are obtained from the output of the MCMC sampling scheme and are shown in Figures \ref{}--\ref{}
%\subsection{Contagion Effects}
\section{Discussion}
Understanding the dynamics of terrorism is a problem at the forefront of concerns at all levels of society, from policy makers to popular discourse the subject of terrorism by its very nature elicits a visceral need for an explanation.   
There has been substantial efforts to map and explore how terrorism spreads in space over time. Research such as \cite{OLoughlin:etal:2010} used data from Wikileaks documents to geocode and map the patterns of violence in Afghanistan from 2004 through 2009, \cite{Linke:etal:2012} conducted a similar study of activity in Iraq finding that there appeared to be a reciprocal relationship between insurgents and counter insurgency activity, while  \cite{Schutte:Weidmann:2011} takes in a broader set of contexts, and observed that the spread of violence mainly manifests as either an increase in geographical area, or as a displacement in space.  

While these papers provide useful insights and clarity to the discussion of events by focus on exploratory techniques to describe the patterns, the focus of this paper is to use a statistical modelling approach to identify distinct differences in the patterns of proliferation, and make inference on those differences.   
The distinction being that in this paper the terms diffusion and contagion refer to mechanisms of proliferation rather than observable phenomena. 
%\mdp{remove: It is hoped that this goal is demonstrated sufficiently well to make the case for the merit of this approach to analysing terrorism. }

Results from the models presented in this paper provide two immediately evident results: in general, key shifts in the diffusion rate coincide with key events in the historical narrative; that is the notion of precipitants as exogenous factors is evident in the models, and endogenous excitation effects tends to decrease as the number of fatalities increase.  
While the use of Afghanistan, Iraq, and Israel as examples may seem limiting, as the results of these analyses may not be extensible to other countries, the results of applying this model to these data and others not included (Turkey, Indonesia, and Colombia) show that the model can be applied broadly. 
%Iraq and Afghanistan may seem similar, but Figures \ref{afghan_nkill} and \ref{iraq_nkill} show that their effects of fatalities on the degree of excitation or ``contagiousness" of events is strikingly different. 

%Israel is a very different context from Afghanistan and Iraq and yet the model results still provide useful insight.  
%Each of these countries is unique and arguably the model results are not necessarily extensible to other contexts, the model presented here is. 

The results for all three countries agree with the existing narratives.  Both Afghanistan and Iraq were under the rule of repressive regimes prior to invasion by US and Allied forces.  Occupation rule in both countries resulted in increased violence which escalated rapidly after the withdrawal of forces and the ceding of authority for security to local forces. 
The difference between the two is that resistance in Afghanistan came primarily in the form of a reassertion of the deposed regime, the Taliban. In Iraq the regime of Saddam Hussein, including the Iraqi army and his Ba'ath party, were systematically dismantled in an effort to create a democratic secular government, creating a power vacuum and exacerbating sectarian tensions between the Shia majority and Sunni minority, contributing to the violence and unrest. This difference is manifest in the differences in the endogenous effects or volatility as measured by $\delta_s$ in Figures \ref{afghan_nkill} and \ref{iraq_nkill}.  
The volatility in Afghanistan is relatively constant, with little decrease or change with the number of fatalities, this indicates that there is likely no inhibitory or ``blow-back'' effects for large scale events. 
%\mdp{[This sentence is not so clear to me with the negations:] 
Coupled with the information that the fatalities and frequency of events both increase in time (as the Taliban consolidates and rebuilds post-invasion) the stable volatility also indicates evidence that the execution of large-scale events is not an excessive drain on resources or capacity to attack.  
%}
In Iraq the volatility shows more variation, as $\delta_s$ decreases with the number of fatalities.  The volatility in excess of $1$ for events with $1$ or $0$ fatalities reflects that numerous groups or individuals were responsible for the events.  The decrease in volatility with fatalities reflects either an inhibitory effect for large scale events, or the limited resources of groups or individuals to act.  In either case the distinction between these patterns is informative. 

The pattern in Israel reflects a relatively stable situation, where terrorist activity ebbs and flows.  A precarious peace process exists between Israeli and Palestinian claims over the region. Describing the situation as complex is an understatement.  In reality, the Israelis and Palestinians share not only claims of sovereignty over the region, but also cultural and economic ties.   The ebb and flow of terrorist activity is a part of daily life and exists almost as an extension of political discourse. Increases in activity often result in new rounds of negotiation, or at least proportional responses rather than all-out offences.  This is reflected in both the variation of the diffusion rate $\mu_t^d$, which does not reach the levels of intensity as in Afghanistan and Iraq, and the volatility which is consistently less than $1$, indicating a stationary process, and the decrease in volatility as a function of fatalities.  Given the relatively low diffusion rate, and the historical context, where violence is often used to exert political pressure, it is more likely that the decrease in volatility as fatalities increase is likely due to an inhibitory effect.  

The classification of mechanisms for the proliferation of terrorist activity into the categories of diffusion and contagion provides a useful framework for analysing activity.  
The intent in developing these mathematical models was to create a means of both analysing and classifying data, but also to test and measure the effectiveness of counter-measures.   
The correspondence between the model results and the existing narratives for the countries analysed validate the use of these models, both conceptually and mathematically. % for that purpose.  
The validation of these models also has important implications for counter-measures.  
First is the distinction between endogenous and exogenous effects, and their roles in the contagion and diffusion processes.  
Second is the implication that the dominant mechanism should guide counter-measures, both counter-terrorism efforts and counter-radicalisation efforts.  
The contagion process is characterised by small-scale dynamics and is governed by endogenous effects, i.e.\ it is the characteristics of these events and immediate or tactical counter-measures that have potential to  
%\mdp{?potential to?} 
influence the contagion process.  
The diffusion process is associated with exogenous factors, or large-scale socio-economic and political factors, these are factors that take a different, more strategic set of counter-measures, in order to effect change.  
Both cases require different approaches, a more tactical approach for addressing factors that effect the contagion process, and a more strategic approach to address factors effecting the diffusion process. If, as is likely, real-world situations are a mixture of both diffusion and contagion then counter-terrorism and counter-radicalisation policies should be shaped to address the two processes and their particular balance in each setting.
%this implies a need for counter-terrorism and counter-radicalisation policies that address these two processes and the particular balance of these processes.\mdp{ not clear: If real-world situations are a mixture of both processes then this implies a need for counter-terrorism and counter-radicalisation policies that address these two processes and the particular balance of these processes.}  
The models here provide a useful tool for analysis to both assess the mechanisms at work in any given context and to measure the effectiveness of enacted counter measures.  
In practice, counter-terrorism and counter-radicalisation activities typically carry the potential of negative consequences, an understanding of the mitigation and prevention options and a method of measuring their effectiveness is important to minimising possible negative effects. 
%Original: In practice, counter-terrorism and counter-radicalisation activities typically have the possibility of negative consequences, an effective measure of both what measures are needed and their effectiveness is important to minimising potential negative effects.  
Thus the utility of the models proposed here is beyond academic and offer substantial benefit to real-world policy makers.

\bibliographystyle{plain}
\bibliography{master_refs}

\begin{thebibliography}{10}

\bibitem{Haaretz:2010}
{U.S. Plans Jeurusalem Talks After Egypt Peace Summit}.
\newblock {\em Haaretz}, 2010.

\bibitem{Bandura:1990}
A.~Bandura.
\newblock {M}echanisms of {M}oral {D}isengagement.
\newblock In W.~Reich, editor, {\em Origins of Terrorism: Psychologies,
  {I}deologies, {T}heologies, {S}tates of {M}ind}, pages 161--191. Cambridge
  University Press, Cambridge, 1990.

\bibitem{Bebber:2008}
Robert~J. Bebber.
\newblock {The Role of Provincial Reconstruction Teams (PRTs) in
  Counterinsurgency Operations: Khost Province, Afghanistan}.
\newblock {\em {Small Wars Journal}}, 2008.

\bibitem{Bennet:2003}
James Bennet.
\newblock {With Cease-Fire Crumbling, Israel Refuses to Suspend Raids Against
  Palestinians}.
\newblock {\em {The New York Times}}, 2003.

\bibitem{Bergen:2010}
Peter Bergen.
\newblock {U.S}. intelligence briefing: {T}aliban increasingly effective, 2010.

\bibitem{loepucl1354681}
A~Braithwaite and SD~Johnson.
\newblock {S}pace-{T}ime {M}odeling of {I}nsurgency and {C}ounterinsurgency in
  {I}raq.
\newblock {\em Journal of Quantitative Criminology}, 28(1):31 -- 48, March
  2012.

\bibitem{Braithwaite:2010}
Alex Braithwaite.
\newblock {R}esiting {I}nfection: {H}ow {S}tate {C}apacity {C}onditions
  {C}onflict {C}ontagion.
\newblock {\em Journal of Peace Research}, 47(3):311--319, 2010.

\bibitem{Brynjar:Skjolberg:2000}
L.~Brynjar and Katja Skj{\o}lberg.
\newblock {W}hy {T}errorism {O}ccurs - {A} {S}urvey of {T}heories and
  {H}ypotheses on the {C}auses of {T}errorism.
\newblock Technical report, Norwegian Defence Research Establishment, 2000.
\newblock http://www.ffi.no/no/rapporter/00-02769.pdf.

\bibitem{Buhaug:Gleditsch:2008}
Halvard Buhaug and Kristian~Skrede Gleditsch.
\newblock {C}ontagion or {C}onfusion? {W}hy {C}onflicts {C}luster in {S}pace.
\newblock {\em International Studies Quarterly}, 58(2):215--233, 2008.

\bibitem{Bush:2007}
George~W. Bush, 2007.

\bibitem{Cliff:First:2013}
Christina Cliff and Andrew First.
\newblock {T}esting for {C}ontagion/{D}iffusion of {T}errorism in {S}tate
  {D}yads.
\newblock {\em Studies in Conflict \& Terrorism}, 36(4):292--314, 2013.

\bibitem{CNN:2004}
CNN.
\newblock Marines, {I}raqis join forces to shut down {F}allujah, 2004.

\bibitem{Coleman:1964}
James~S. Coleman.
\newblock {\em {I}ntroduction to {M}athematical {S}ociology}.
\newblock Free Press, 1964.

\bibitem{Crenshaw:1981}
Martha Crenshaw.
\newblock {T}he {C}auses of {T}errorism.
\newblock {\em Comparative Politics}, 13(4):379--399, 1981.

\bibitem{Crenshaw:1986}
Martha Crenshaw.
\newblock {\em {P}olitical {P}sychology: {C}ontemporary {P}roblems and
  {I}ssues}, chapter The Psychology of Political Terrorism, pages 379--413.
\newblock Jossey-Bass, 1986.
\newblock Reprinted in John T. Jost and Jim Sidanius, eds., Political
  Psychology: Key Readings (Taylor and Francis, 2004).

\bibitem{D/VJ:03}
D.~J. Daley and D.~Vere-Jones.
\newblock {\em {A}n {I}ntroduction to the {T}heory of {P}oint {P}rocesses},
  volume~I.
\newblock Springer-Verlag, New York, NY, 2nd edition, 2003.

\bibitem{Diekmann:1979}
Andreas Diekmann.
\newblock {A} {D}ynamic {S}tochastic {V}ersion of the
  {P}itcher-{H}amblin-{M}iller {M}odel of ``{C}ollective {V}iolence''.
\newblock {\em Journal of Mathematical Sociology}, 6(2):277--282, 1979.

\bibitem{Einstein:1905}
Albert Einstein.
\newblock {\"U}ber die von der molekularkinetischen {T}heorie der {W}{\"a}rme
  geforderte {B}ewegung von in ruhenden {F}l{\"u}ssigkeiten suspendierten
  {T}eilchen.
\newblock {\em Annalen der Physik}, 17(8):549--560, 1905.

\bibitem{Kareem:Hwaida:2013}
Kareem Fahim and Hwaida Saad.
\newblock {A Faceless Teenage Refugee Who Helped Ignite Syria's War}.
\newblock {\em {The New York Times}}, 2013.

\bibitem{Felbab-Brown:2012}
V.~Felbab-Brown.
\newblock {Slip-Sliding on a Yellow Brick Road: Stabilization Efforts in
  Afghanistan}.
\newblock {\em {Stability: International Journal of Security and Development}},
  pages 4--19, 2012.

\bibitem{Feller:1943}
William Feller.
\newblock {O}n a {G}eneral {C}lass of ``{C}ontagious'' {D}istributions.
\newblock {\em Annals of Mathematical Statistics}, 14(4):389--400, 1943.

\bibitem{Fick:1855}
Adolph Fick.
\newblock {\"U}ber {D}iffusion.
\newblock {\em Poggendorff's Annalen}, 94:59--86, 1855.

\bibitem{Flaherty:2007}
A.~Flaherty.
\newblock {Petraeus Talks of Troop Withdrawal}.
\newblock {\em {The Associated Press}}, 2007.

\bibitem{Carlotta:2004}
Carlotta Gall.
\newblock {Taliban Leader Vows Return}.
\newblock {\em {The New York Times}}, 2004.

\bibitem{Goldberg:2000}
Suzanne Goldberg.
\newblock Rioting as {S}haron visits {I}slam holy site.
\newblock {\em {The Guardian}}, 2000.

\bibitem{Greenwood:Yule:1920}
Major Greenwood and G.~Udny Yule.
\newblock {A}n {I}nquiry into the {N}ature of {F}requency {D}istributions
  {R}epresentative of {M}ultiple {H}appenings with {P}articular {R}eference to
  the {O}ccurrence of {M}ultiple {A}ttacks of {D}isease or of {R}epeated
  {A}ccidents.
\newblock {\em Journal of the Royal Statistical Society}, 83(2):pp. 255--279,
  1920.

\bibitem{Gurr:1993}
Ted~Robert Gurr.
\newblock {W}hy {M}inorities {R}ebel: {A} {G}lobal {A}nalysis of {C}ommunal
  {M}obilization and {C}onflict since 1945.
\newblock {\em International Political Science Review}, 14(2):161--201, 1993.

\bibitem{Hamilton:Hamilton:1981}
James~D. Hamilton and Lawrence~C. Hamilton.
\newblock {M}odels of social contagion.
\newblock {\em The Journal of Mathematical Sociology}, 8(1):133--160, 1981.

\bibitem{Hamilton:Hamilton:1983}
Lawrence~C. Hamilton and James~D. Hamilton.
\newblock {D}ynamics of {T}errorism.
\newblock {\em International Studies Quarterly}, 27(1):39--54, 1983.

\bibitem{Hashim:2005}
Ahmed Hashim.
\newblock {\em Insurgency and Counter-insurgency in Iraq}.
\newblock Cornell University Press, 2005.

\bibitem{hawkes71b}
Alan~G. Hawkes.
\newblock {P}oint spectra of some mutually exciting point processes.
\newblock {\em Journal of the Royal Statistical Society. Series B},
  33(3):438--443, 1971.

\bibitem{Hawkes:1971a}
Alan~G. Hawkes.
\newblock {S}pectra of {S}ome {S}elf-{E}xciting and {M}utually {E}xciting
  {P}oint {P}rocesses.
\newblock {\em Biometrika}, 58(1):83--90, 1971.

\bibitem{hawkes&oakes:1974}
Alan~G. Hawkes and David Oakes.
\newblock {A} {C}luster {P}rocess {R}epresentation of a {S}elf-{E}xciting
  {P}rocess.
\newblock {\em Journal of Applied Probability}, 11(3):pp. 493--503, 1974.

\bibitem{Heyman:Micklous:1980}
Edward Heyman and Edward Mickolus.
\newblock {O}bservations on ``{W}hy {V}iolence {S}preads''.
\newblock {\em International Studies Quarterly}, 24(2):299--305, 1980.

\bibitem{Hill:Rothchild:1986}
Stuart Hill and Donald Rothchild.
\newblock {T}he {C}ontagion of {P}olitical {C}onflict in {A}frica and the
  {W}orld.
\newblock {\em Journal of Conflict Resolution}, 30(4):23--35, 1986.

\bibitem{Hodge:2013}
Nathan Hodge.
\newblock {Blast Mars Day of Security Handover in Kabul}.
\newblock {\em {The Wall Street Journal}}, 2013.

\bibitem{Hoffman:2006}
Bruce Hoffman.
\newblock {\em {I}nside {T}errorism}.
\newblock Columbia University Press, 2006.

\bibitem{Holden:1986}
R.T. Holden.
\newblock {T}he {C}ontagiousness of {A}ircraft {H}ijacking.
\newblock {\em American Journal of Sociology}, 91(91):874--904, 1986.

\bibitem{Hopper:1950}
Rex~D. Hopper.
\newblock {T}he {R}evolutionary {P}rocess: {A} {F}rame of {R}eference for the
  {S}tudy of {R}evolutionary {M}ovements.
\newblock {\em Social Forces}, 28(3):pp. 270--279, 1950.

\bibitem{Hutchinson:1972}
M.~C. Hutchinson.
\newblock {T}he concept of revolutionary terrorism.
\newblock {\em The Journal of Conflict Resolution}, 16(3):383--396, 1972.

\bibitem{Johnson:etal:2014}
Kay Johnson, Raissa Kasolowsky, Michael Perry, and Kevin Liffey.
\newblock Britain ends combat role in afghanistan, last us marines hand over
  base, 2014.

\bibitem{Keegan:2005}
John Keegan.
\newblock {\em The Iraq Awards}.
\newblock Vintage Books, 2005.

\bibitem{Knights:2011}
Michael Knights.
\newblock {The JRTN Movement and Iraq's Next Insurgency}.
\newblock {\em CTC Sentinel}, 4:1--6, 2011.

\bibitem{Lafree:etal:2009}
G.~LaFree, L.~Dugan, and R.~Korte.
\newblock {T}he {I}mpact of {B}ritish {C}ounter {T}errorist {S}trategies on
  {P}olitical {V}iolence in {N}orthern {I}reland: {C}omparing {D}eterrence and
  {B}acklash {M}odels.
\newblock {\em Criminology}, 47:17--45, 2009.

\bibitem{Landler:Cooper:2011}
Mark Landler and Helene Cooper.
\newblock {Obama Will Speed Pullout From War in Afghanistan}.
\newblock {\em {The New York Times}}, 2011.

\bibitem{Lang:Brezger:2001}
Stefan Lang and Andreas Brezger.
\newblock {Bayesian P-Splines}.
\newblock techreport, University of Munich, 2001.

\bibitem{LeBon:1896}
Gustav {Le Bon}.
\newblock {\em {T}he {C}rowd: {A} {S}tudy of the {P}opular {M}ind}.
\newblock Macmillian, 1896.

\bibitem{Li:Thompson:1975}
R.~P. Li and W.~R. Thompson.
\newblock {T}he ``{C}oup {C}ontagion'' {H}ypothesis.
\newblock {\em The Journal of Conflict Resolution}, 19:63--88, 1975.

\bibitem{Linke:etal:2012}
Andrew~M. Linke, Frank~D.W. Witmer, and John O'Loughlin.
\newblock {Space-time Granger analysis of the war in Iraq: A study of coalition
  and insurgent action-reaction}.
\newblock {\em International Interactions}, 38(4):402--425, 2012.

\bibitem{MERMI:2013}
MEMRI.
\newblock {ISI Confirms That Jabhat Al-Nusra Is Its Extension in Syria,
  Declares 'Islamic State of Iraq And Al-Sham' As New Name of Merged Group}.
\newblock {\em Middle East Media Research Institute}, 2013.

\bibitem{Midlarsky:1978}
M.~I. Midlarsky.
\newblock {A}nalyzing diffusion and contagion effects: {T}he urban disorders of
  the 1960s.
\newblock {\em The American Political Science Review}, 72:996--1008, 1978.

\bibitem{Midlarsky:etal:1980}
M.~I Midlarsky, M.~Crenshaw, and F.~Yoshida.
\newblock {W}hy {V}iolence {S}preads: {T}he {C}ontagion of {I}nternational
  {T}errorism.
\newblock {\em International Studies Quarterly}, 24:341--365, 1980.

\bibitem{Midlarsky:1970}
Manus~I. Midlarsky.
\newblock {Mathematical Models of Instability and a Theory of Diffusion}.
\newblock {\em International Studies Quarterly}, 14(1), March 1970.

\bibitem{Myre:Erlanger:2006}
Greg Myre and Steven Erlanger.
\newblock {I}sraelis {E}nter {L}ebanon {A}fter {A}ttacks.
\newblock {\em {The New York Times}}, 2006.

\bibitem{GTD:2017}
{National Consortium for the Study of Terrorism and Responses to Terrorism
  (START)}.

\bibitem{GTD:2012}
{National Consortium for the Study of Terrorism and Responses to Terrorism
  (START)}.
\newblock {G}lobal {T}errorism {D}atabase [{D}ata file], June 2015.
\newblock Retrieved from http://www.start.umd.edu/gtd.

\bibitem{Neyman:1939}
J.~Neyman.
\newblock {O}n a {N}ew {C}lass of ``{C}ontagious'' {D}istributions,
  {A}pplicable in {E}ntomology and {B}acteriology.
\newblock {\em Annals of Mathematical Statistics}, 10(1):35--57, 03 1939.

\bibitem{OBryant:Waterhouse:2008}
JoAnne O'Bryant and Michael Waterhouse.
\newblock {U.S. Forces in Afghanistan}.
\newblock {\em {CRS} {R}eport for {C}ongress}, 2008.

\bibitem{OHanlon:2010}
Michael O'Hanlon.
\newblock {Staying Power: The U.S. Mission in Afghanistan Beyond 2011}.

\bibitem{OLoughlin:etal:2010}
John O'Loughlin, Frank~DW Witmer, and Andrew~M Linke.
\newblock {The Afghanistan-Pakistan wars, 2008-2009: Micro-geographies,
  conflict diffusion, and clusters of violence}.
\newblock {\em Eurasian Geography and Economics}, 51(4):437--471, 2010.

\bibitem{Onsager:1931}
Lars Onsager.
\newblock {R}eciprocal {R}elations in {I}rreversible {P}rocesses. {I}.
\newblock {\em Physical Review}, 37:405--426, Feb 1931.

\bibitem{Philibert:2005}
Jean Philibert.
\newblock {O}ne and a {H}alf {C}entury of {D}iffusion: {F}ick, {E}instein,
  before and beyond.
\newblock {\em Diffusion Fundamentals}, 2:1--10, 2005.

\bibitem{Pitcher:etal:1978}
Brian~L. Pitcher, Robert~L. Hamblin, and Jerry L.~L. Miller.
\newblock {T}he {D}iffusion of {C}ollective {V}iolence.
\newblock {\em American Sociological Review}, 43(1):23--35, 1978.

\bibitem{Porter:White:2012}
Michael~D. Porter and Gentry White.
\newblock {S}elf-exciting {H}urdle {M}odels for {T}errorist {A}ctivity.
\newblock {\em Annals of Applied Statistics}, 6(1):106--124, 2012.

\bibitem{AP:2007}
Associated Press.
\newblock {Iraq Bill Demands U.S. Troop Withdraw}, 2007.

\bibitem{Rashid:2000}
Ahmed Rashid.
\newblock {\em Taliban}.
\newblock Yale University Press, 2000.

\bibitem{Reuters:2010}
Reuters.
\newblock Karzai holds peace talks with insurgents, 2010.

\bibitem{Reuters:2012}
Reuters.
\newblock {NATO} sets ``irreversible'' but risky course to end {A}fghan war,
  2012.

\bibitem{Ricks:2007}
Thomas~E. Ricks.
\newblock {\em Fiasco}.
\newblock Penguin Books, 2007.

\bibitem{Robert:Casella:2004}
Christian~P. Robert and George Casella.
\newblock {\em {M}onte {C}arlo {S}tatistical {M}ethods}.
\newblock Springer, 2 edition, 2004.

\bibitem{Rouder:etal:2003}
J.~N. Rouder, D.~Sun, P.~L. Speckman, J.~Lu, and D.~Zhou.
\newblock {A} hierarchical {B}ayesian statistical framework for skewed
  variables with an application to response time distributions.
\newblock {\em Psychometrika}, 68(4):589--606, 2003.

\bibitem{Ruthven:2016}
Malise Ruthven.
\newblock {How to Understand ISIS}.
\newblock {\em {The New York Review of Books}}, 63(11), 2016.

\bibitem{Schmid:Jongman:2005}
Alex Schmid and A.~Jongman.
\newblock {\em {P}olitical {T}errorism}.
\newblock Transaction Publishers, 2005.

\bibitem{Schutte:Weidmann:2011}
Sebastian Schutte and Nils~B. Weidmann.
\newblock {Diffusion Patterns in Civil Wars}.
\newblock {\em Political Geography}, 30:143--152, 2011.

\bibitem{Shadid:2010}
Anthony Shadid.
\newblock {Iraqi Insurgent Group Names New Leaders}.
\newblock {\em {The New York Times}}, 2010.

\bibitem{Smith:1996}
Robert~L. Smith.
\newblock The {H}it-and-{R}un {S}ampler: {A} {G}lobally {R}eaching {M}arkov
  {C}hain {S}ampler for {G}enerating {A}rbitrary {M}ultivariate
  {D}istributions.
\newblock In J.~M. Charnes, D.~J. Morrice, D.~T. Brunner, , and J.~J. Swain,
  editors, {\em Proceedings of the 1996 Winter Simulation Conference}, 1996.

\bibitem{Syal:2014}
Ryan Syal.
\newblock {UK} troops hand over {C}amp {B}astion to {A}fghan forces, ending
  13-year campaign.
\newblock {\em {The Guardian}}, 2014.

\bibitem{Telesca:Lovallo:2006}
L.~Telesca and M.~Lovallo.
\newblock {A}re global terrorism attacks time-correlated?
\newblock {\em Physica A}, 362:480--484, 2006.

\bibitem{Teorell:1936}
Torsten Teorell.
\newblock {S}tudies on the ``{D}iffusion {E}ffect'' {U}pon {I}conic
  {D}istribution {I}. {S}ome {T}heoretical {C}onsiderations.
\newblock {\em Proceedings of the National Academy of Sciences},
  21(3):152--161, 1936.

\bibitem{Thorton:1964}
T.~P. Thornton.
\newblock {\em {I}nternal {W}ar}, chapter Terror as a weapon of political
  aggitation, pages 71--99.
\newblock Free Press, New York, 1964.

\bibitem{Thrall:2014}
Nathan Thrall.
\newblock Hamas's {C}hances.
\newblock {\em London Review of Books}, 2014.

\bibitem{Tohid:Scott:2003}
Owias Tohid and Scott Baldauf.
\newblock Taliban appears to be regrouped and well-funded.
\newblock {\em Christian Science Monitor}, 2003.

\bibitem{Townsley:2008}
M.~Townsley.
\newblock {V}isualising {S}pace {T}ime {P}atterns in {C}rime: {T}he {H}otspot
  {P}lot.
\newblock {\em Crime Patterns and Analysis}, 1:61--74, 2008.

\bibitem{VandenBrook:2011}
Tom Vanden~Brook.
\newblock {Raids Have Taken Out 900 Taliban Leaders}.
\newblock {\em USA Today}, 2011.

\bibitem{Weinmann:etal:1988}
Gabriel Weimann and Hans-Bernd Brosius.
\newblock {T}he predictability of international terrorism: {A} time-series
  analysis.
\newblock {\em Terrorism}, 11(6):491--502, 1988.

\bibitem{White:Porter:2013}
Gentry White and Michael~D. Porter.
\newblock {GPU} {A}ccelerated {MCMC} for {M}odelling {T}errorist {A}ctivity.
\newblock {\em Computational Statistics and Data Analysis}, 71(C):643--651,
  2013.

\bibitem{White:etal:2012b}
Gentry White, Michael~D. Porter, and Lorraine Mazerolle.
\newblock {T}errorism {R}isk, {R}esilience and {V}olatility:{A} {C}omparison of
  {T}errorism in {T}hree {S}outheast {A}sian {C}ountries.
\newblock {\em Journal of Quantitative Criminology}, 29(2):295--320, 2013.
\newblock DOI 10.1007/s10940-012-9181-y.

\end{thebibliography}

\end{document}